\documentclass[apj]{emulateapj}
\usepackage{apjfonts}
\usepackage{amsbsy}

\def\wp{$w_p(r_p)$}
\def\hmpc{$h^{-1}$Mpc}
\def\hkpc{$h^{-1}$kpc}

\def\hmsol{$h^{-1}$M$_\odot$}

\def\navg{\langle N\rangle_M}

\def\nsat{\langle N_{\rm sat}\rangle_M}
\def\ncen{\langle N_{\rm cen}\rangle_M}
\def\mmin{M_{\rm min}}

\def\msat{M_{\rm sat}}

\def\asat{\alpha_{\rm sat}}

\def\om{\Omega_m}
\def\oml{\Omega_\Lambda}
\def\omb{\Omega_b}
\def\s8{\sigma_8}

\def\lcdm{$\Lambda$CDM}

\def\x2{$\chi^2$}
\def\hmsol{$h^{-1}\,$M$_\odot$}

\def\NNm1{\langle N(N-1) \rangle}

\def\slogm{\sigma_{{\rm log}M}}
\def\m_star{M_\ast}

\def\lcdm{$\Lambda$CDM}
\def\slogm{\sigma_{{\rm log}M}}
\def\om{\Omega_m}

\def\omb{\Omega_b}
\def\s8{\sigma_8}

\def\hmpc{$h^{-1}\,$Mpc}
\def\hkpc{$h^{-1}\,$kpc}

\def\x2{$\chi^2$}
\def\hmsol{$h^{-1}\,$M$_\odot$}

\def\wp{$w_p(r_p)$}

\def\mmin{M_{\rm min}}

\def\sigmaM{\sigma_{\log M}}
\def\navg{\langle N\rangle_M}

\def\nsat{\langle N_{\mbox{\scriptsize sat}}\rangle_M}

\def\ncen{\langle N_{\mbox{\scriptsize cen}}\rangle_M}

\def\NNm1{\langle N(N-1) \rangle}

\def\sigmaM{\sigma_{\log M}}

\def\p0{P_0(r)}

\def\msat{M_{\rm sat}}

\def\m12{M_{12}}

\def\m12{M_{12}}

\def\rs0{\hat{R}_{\rm sh}^0}
\def\mg2{Mg\,II}

\def\mcrit{M_{\rm crit}}

\def\wth{w(\theta)}

\def\arcsec{^{\prime\prime}}
\def\wp{w_p(r_p)}
\def\xisp{\xi(r_p,\pi)}
\def\frsat{f_{\rm Rsat}}
\def\frcen{f_{\rm Rcen}}
\def\frmax{f_{\rm Rmax}}
\def\mredmblue{M_{\rm red}/M_{\rm blue}}
\def\nblue{n_{\rm blue}}
\def\nred{n_{\rm red}}
\def\nprev{\bar{n}_{\rm prev}}
\def\nsatdens{\bar{n}_{\rm sat}}
\def\fq{f_{\rm Q}}
\def\tq{t_{\rm Q}}
\def\msub{M_{\rm sub}}
\def\mhost{M_{\rm host}}
\def\tsat{t_{\rm sat}}
\def\mgal{M_{\rm gal}}
\def\fm{f_{\rm merge}}
\def\msub{M_{\rm sub}}

\begin{document}

\title{What Does Clustering Tell Us About the Buildup of the Red Sequence?}

\author{ Jeremy L. Tinker$^1$ and Andrew R. Wetzel$^2$ }
\affil{$^1$Berkeley Center for Cosmological Physics, University of California-Berkeley\\
  $^2$ Department of Astronomy, University of California-Berkeley}

\begin{abstract}

  We analyze the clustering of red and blue galaxies from four samples
  spanning a redshift range of $0.4<z<2.0$ to test the various
  scenarios by which galaxies evolve onto the red sequence. The data
  are taken from the UKIDSS Ultra Deep Survey, DEEP2, and
  COMBO-17. The use of clustering allows us to determine what fraction
  of the red sequence is made up of central galaxies and satellite
  galaxies. At all redshifts, including $z=0$, the data are consistent
  with $\sim 60\%$ of satellite galaxies being red or quenched,
  implying that $\sim 1/3$ of the red sequence is comprised of
  satellite galaxies. More than three-fourths of red satellite
  galaxies were moved to the red sequence after they were accreted
  onto a larger halo. The constant fraction of satellite galaxies
  that are red yields a quenching time for satellite galaxies that
  depends on redshift in the same way as halo dynamical times;
  $\tq\sim (1+z)^{-1.5}$. In three of the four samples, the data favor
  a model in which red central galaxies are a random sample of all
  central galaxies; there is no preferred halo mass scale at which
  galaxies make the transition from star-forming to red and dead. The
  large errors on the fourth sample inhibit any
  conclusions. Theoretical models in which star formation is quenched
  above a critical halo mass are excluded by these data. A scenario in
  which mergers create red central galaxies imparts a weaker
  correlation between halo mass and central galaxy color, but even the
  merger scenario creates tension with red galaxy clustering at
  redshifts above 0.5. These results suggest that the mechanism by
  which central galaxies become red evolves from $z=0.5$ to $z=0$.

\end{abstract}


\section{Introduction}

In the redshift-zero universe we observe a distinct bimodality in the
distribution of galaxies; the two populations are comprised of blue
star-forming objects and red passively evolving objects
(\citealt{strateva_etal:01, blanton_etal:03cmd, kauffmann_etal:03b,
  madgwick_etal:03}). Bimodality is clearly seen at $z\sim 1$
(\citealt{bell_etal:04, cooper_etal:06, willmer_etal:06}). Recent
results detect this bimodality out to $z=2$ and suggest that the red
sequence exists at even higher redshifts (\citealt{kriek_etal:08,
  williams_etal:09}, hereafter W09). How this bimodality was created
and has evolved is an outstanding problem in galaxy evolution and the
history of star formation in the universe. In this paper we use the
clustering of several bimodal galaxy samples spanning $0.4<z<2.0$ to
test the various physical processes that may halt star formation
within a galaxy and initiate its transition onto the red sequence.

There are three dominant physical mechanisms by which galaxies can
migrate from the blue cloud to the red sequence: (1) Major mergers can
rapidly exhaust the available gas within two galaxies in a burst of
star formation, resulting in a passively evolving elliptical galaxy
(e.g., \citealt{toomre_toomre:72, negroponte_white:83,
  mihos_hernquist:96, springel:00}); this idea has been developed in a
more statistical fashion by \cite{hopkins_etal:08a,hopkins_etal:08b}.
Since the galaxy merger rate is expected to increase with mass, this
scenario implies a larger red fraction at higher halo mass.  (2)
Galaxies can be accreted onto a group-sized or cluster-sized dark
matter halo; in this scenario, the new satellite halo (which we will
refer to as a subhalo) is no longer able to accrete gas, and the gas
already contained within the subhalo is subjected to both tidal forces
and ram-pressure stripping. Thus star formation is predicted to be
attenuated in satellite galaxies (e.g., \citealt{gunn_gott:72,
  abadi_etal:99, wang_etal:07, vdb_etal:08, kimm_etal:09}). This
effect is observed in the color distributions of clusters of galaxies
(e.g., \citealt{butcher_oemler:78, butcher_oemler:84,
  hansen_etal:09}). (3) There is a halo mass scale above which star
formation is significantly attenuated, either by virial shocks and
related processes (\citealt{birnboim_dekel:03, keres_etal:05,
  keres_etal:08, dekel_birnboim:06, cattaneo_etal:06,
  dekel_birnboim:08}) or through feedback by active galactic nuclei,
which is strongly correlated with halo mass (\citealt{croton_etal:06a,
  bower_etal:06, delucia_blaizot:07, somerville_etal:08}). Many of
these studies give rise to a critical mass scale of $\sim 10^{12}$
\hmsol, above which shock heating within a halo is efficient and gas
accretion enters the ``hot mode'' phase. Outstanding questions remain:
When do these processes begin and how long do they persist?  Which
dominate at high redshift? How is the red sequence built over time?

In this paper, we analyze previously published clustering measurements
of red (quiescent) and blue (star-forming) galaxies for three
different surveys at three distinct redshifts. W09 presented
clustering results from galaxies in the UKIDSS Ultra Deep Survey
(UDS). At $1<z<2$, they demonstrate a clear bimodal sequence in the
galaxy population between star-forming and quiescent galaxies, with
the quiescent systems being significantly more clustered. At
$0.7<z<1.2$, \cite{coil_etal:08} presented color-dependent clustering
measurements from the Deep Extragalactic Evolution Probe (DEEP2)
spectroscopic survey. Once again, the red galaxies showed stronger
clustering at all scales than the blue galaxies for any magnitude
threshold. Using the COMBO-17 photometric survey,
\cite{phleps_etal:06} (hereafter P06) measured the clustering of blue
and red galaxies at $0.4<z<0.8$, with similar results.

Our analysis of this clustering utilizes the halo occupation
distribution (HOD; see, e.g., \citealt{peacock_smith:00, seljak:00,
  roman_etal:01, cooray_sheth:02, berlind_weinberg:02} for early works,
and \citealt{zheng_etal:07, vdb_etal:07, tinker_etal:09_drg} for
examples of more recent implementations of the framework). The HOD
specifies the connection between galaxies and halos on a statistical
basis, a connection that can be constrained through analysis of the
two point correlation function $\xi(r)$. 

Halo occupation analysis of bicolor clustering provides a test of many
of the scenarios described above, since each one makes a distinct
prediction for how red central galaxies occupy their halos and how
they cluster.  This enables us to address many questions related to
the buildup of the red sequence.  What fraction of the red sequence is
created through satellite accretion?  What is the quenching timescale
for this process? At what halo mass scale---if any---do central
galaxies transition from the blue cloud to the red sequence?

Unless otherwise stated, all calculations adopt a flat \lcdm\
cosmology consistent with the latest constraints from CMB anisotropies
(\citealt{dunkley_etal:08}). Our cosmological parameter set is
$(\om,\s8,h,n_s,\omb) = (0.25, 0.8, 0.7, 0.95, 0.045)$. All distances
are in comoving units, and all quoted magnitudes assume
$h=1$. Throughout this paper, we will use the terms ``red'' and
``blue'' interchangeably with ``quiescent/passive'' and ``star
forming.''

\begin{figure}
\epsscale{1.0} 
\plotone{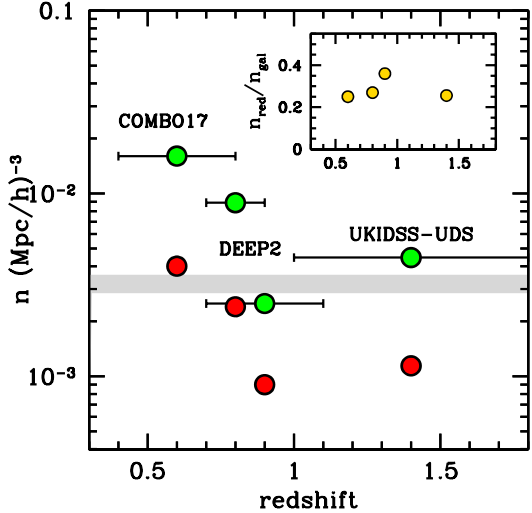}
\caption{ \label{number_densities} Green circles show the number densities of all
  (red+blue) galaxies in each sample as a function of their mean
  redshift. The horizontal error bars show the redshift range of each
  sample. The red circles show the number density of the red subsample
  in each survey. The inset panel shows the fraction of galaxies that
  are red in each sample, which, with the exception of the bright
  DEEP2 sample, is roughly 25\%. The gray shaded band is the number
  density of galaxies brighter than $L_\ast$ at $z=0$ from the SDSS
  $r$-band luminosity function (\cite{blanton_etal:03}).  }
\end{figure}

\begin{figure}
\epsscale{1.0} 
\plotone{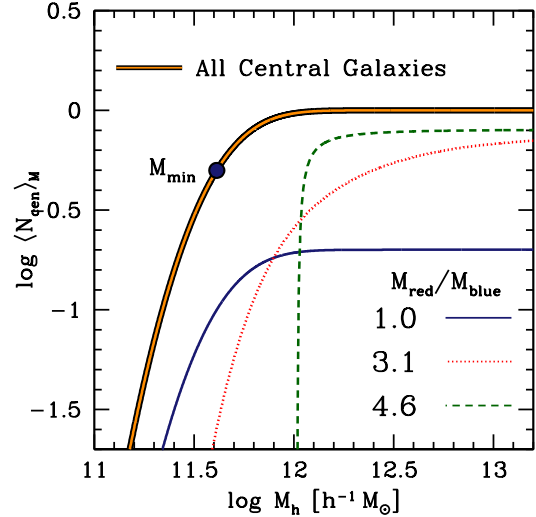}
\caption{ \label{ncen_basic} Examples of various central occupation
  functions for full samples and red galaxy subsamples. The thick
  curve shows a typical $\ncen$ for central galaxies that matches the
  UDS sample. The three thin curves show various $\frcen$ models that
  all have the number of red central galaxies, 20\% of the number of
  all central galaxies. The solid curve is a model in which
  $\frcen=0.2$ at all masses, yielding $\mredmblue=1$. The dotted
  curve shows a model with a broad, smooth transition from halos
  hosting entirely blue central galaxies at low masses to
  predominantly red central galaxies at the group and cluster scale,
  yielding $\mredmblue=3.1$. This model is similar to a scenario in
  which central galaxies become red through major mergers. The dashed
  curve shows a model that contains a critical mass scale at
  $M=10^{12}$ \hmsol, above which 80\% of central galaxies are red,
  yielding $\mredmblue=4.6$. This model is similar to a scenario as
  outlined by \cite{dekel_birnboim:06}.}
\end{figure}

\begin{figure*}
\epsscale{1.2} 
\plotone{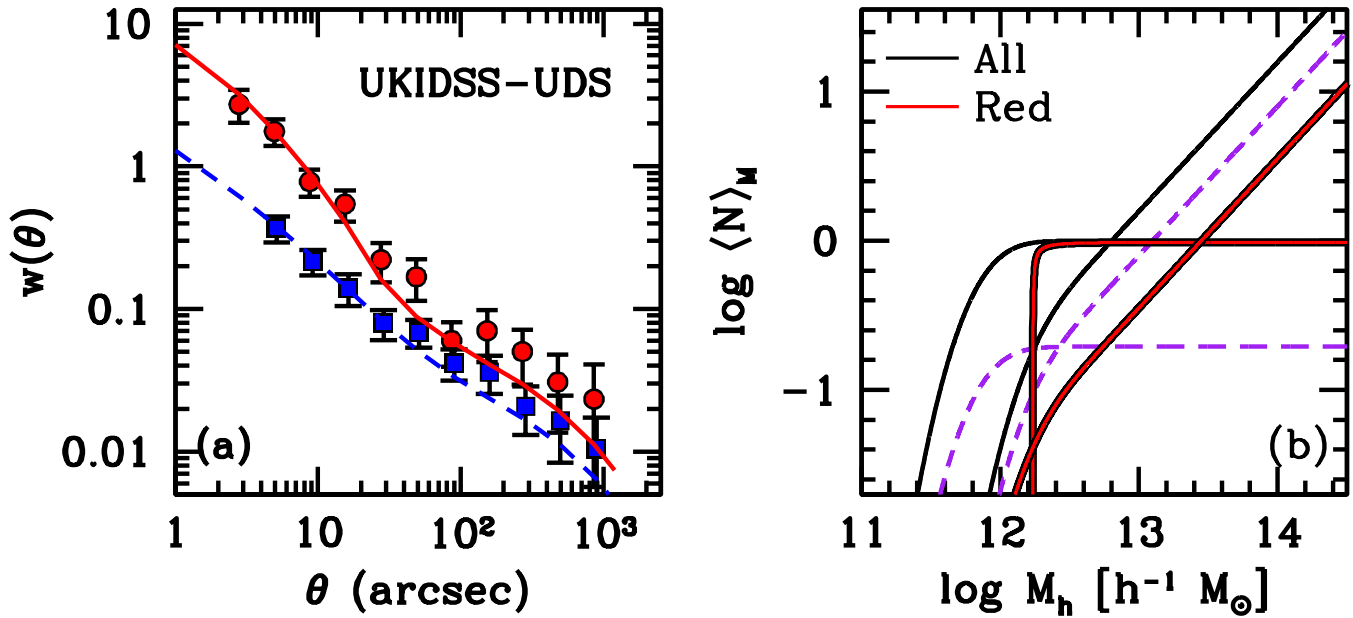}
\vspace{-10.0cm}
\caption{ \label{wtheta_uds} Panel (a): Clustering measurements from
  \cite{williams_etal:09} for quiescent and star-forming galaxies in
  the Ultra Deep Survey of UKIDSS within the redshift range
  $1<z<2$. Circles represent the quiescent subsample, while squares
  represent the star-forming galaxies. The solid and dotted curves
  show the best-fit HOD model for quiescent and star-forming galaxies,
  respectively. Panel (b): Halo occupation functions from the best-fit
  model. The thin solid curves indicate the HOD for {\it all} galaxies
  in the sample; the horizontal curve with a Gaussian cutoff at
  $M\lesssim 10^{11.8}$ \hmsol\ represents all central galaxies, while
  the power-law curve represents satellite galaxies. The thick, red
  curves indicate the fraction of these galaxies that are quiescent
  (red). The dashed curves represent a model in which $\frcen=0.22$
  independent of halo mass. This model is within the $2\sigma$ lower
  limit on the HOD from the clustering data.}
\end{figure*}

\section{Data and Errors}

\subsection{Clustering in the UDS}

The highest redshift clustering data comes from W09, in which the
angular clustering, $\wth$, of star-forming and quiescent galaxies
were measured. The galaxies in the UDS are selected from
observed-frame $K$-band photometry, complete down to $K<22.4$. The
sample analyzed here is a photometric redshift cut at $1<z<2$, with a
median redshift of $z=1.4$. W09 determined the rest-from $UVJ$ colors
of the galaxies in a UDS fields. Within this color-color diagram, two
distinct populations are seen out to $z=2$, with one population having
star formation rates on average an order of magnitude higher than the
other. W09 found that the clustering amplitude of red galaxies is
nearly a factor of two higher than that of blue galaxies at large
scales.''

The errors presented in W09 are bootstrap errors only, accounting for
shot noise but without any estimation of sample variance. To place
more robust errors on these data we use a high-resolution N-body
simulation performed by Martin White using his TPM code
(\citealt{white:02}). The simulation volume is 720 \hmpc\ per side
with 1500$^3$ particles, yielding a mass resolution of $7.67\times
10^9$ \hmsol\ per particle.  The cosmology of the simulation is the
same as that assumed throughout this paper. Using the best-fit HOD
model (utilizing W09 errors only; see in \S 3 for details on HOD
fitting), the halos in the simulation were populated with red and blue
galaxies matching the best-fit occupation functions of each. Although
the photometric redshift cut in the W08 sample is $1<z<2$, the
redshift distribution, $N(z)$ falls off rapidly after $z=1.6$, and the
fraction of galaxies at $z>1.7$ is small.  The box length itself is
not long enough to cover the comoving distance from $1<z<1.7$ (roughly
1 $h^{-1}$Gpc). To circumvent this problem, we reflected the box
around the $x$ and $y$ axes, creating a volume 4 times larger, then
rotated the resulting shape about the $z$ axis by an angle of
45$^\circ$. Within this volume there is a cuboid with $xyz$ dimensions
$720/\sqrt{2}\times 720\sqrt{2}\times 720$ that fully utilizes the
original volume of the simulation with no double-counting of any
structure within the box (here the $y$ axis is the line of sight). The
extra box length comes at the expense of the one other dimension, but
the resulting volume has sufficient comoving depth to model the W09
data within $1<z<1.8$.

Given the angular area of the the UDS, $0.7$ deg$^2$, 140 independent
lines of sight could be obtained from the simulation. In each line of
sight, we calculate $\wth$ with the \cite{landy_szalay:93} estimator,
as in done in W09. Ideally, a full error analysis should include the
entire covariance matrix of both the red and blue clustering
measurements, which, since they are measured from the same patch of
sky, will be correlated. The 140 realizations are not sufficient to
obtain a reliable estimate of the covariance, so throughout this paper
we use diagonal errors only on the W08 data. This renders the
parameter constraints obtained somewhat suspect. To aide in the
constraining power of these data, and to help make up for the lack of
an estimate of the covariance between the red and blue galaxies, we
also use the relative bias of red and blue clustering. This quantity
is less effected by sample variance because the amplitudes of red and
blue galaxies in a given patch of sky will move in concert if the
overall clustering of that patch varies from the universal mean
(\citealt{seljak_etal:09}). Taking the ratio of $\wth$ at each of the
four data points at $\theta>100\arcsec$, where the data are firmly in
the two-halo regime, the relative bias $b_{\rm red}/b_{\rm
  blue}=1.42$. Using the same approach on $\wth$ from the mocks, the
variance of $b_{\rm red}/b_{\rm blue}$ is 11\%. We also incorporate
the error in the number density of each subsample, which we estimate
from the variance of the mocks. These errors are 12\% for the red
galaxies and 14\% for the blue galaxies.

Finally, to model $\wth$, information on the redshift distribution,
$N(z)$, is required.  The photometric redshift distribution is close
to a top-hat function from $1<z<1.8$, but due to uncertainties in the
photometric redshifts, the true $N(z)$ is somewhat broader. An
estimate of the underlying $N(z)$ using the errors in the photo-z's,
kindly provided by R. Williams, is used in all analytic
calculations. The methodology for obtaining the estimated $N(z)$ is
described in W09 and \cite{quadri_etal:08}.

\subsection{Clustering Measurements from the DEEP2 Survey}

Measurements of the clustering of blue and red galaxies at $z\sim 0.8$
in the DEEP2 survey have been published by \cite{coil_etal:08}. Being
a spectroscopic redshift survey, DEEP2 is superior to the UDS in that
the true redshift-space correlation function can be estimated. This
quantity, referred to as $\xisp$, where $r_p$ is the projected
separation of galaxy pairs and $\pi$ is the line-of-sight separation,
can be integrated along $\pi$ to obtain a measurement of the projected
galaxy correlation function, $\wp$. The use of a projected quantity
ameliorates the effect of redshift-space distortions on both large
scales from coherent infall into overdense regions (e.g.,
\citealt{kaiser:87, fisher:95, hamilton:98, roman:04}) and small
scales from non-linearities and virial motions of satellite galaxies
(e.g., \citealt{davis_peebles:83}). The quantity $\wp$ is only free of
redshift space effects if the projection is over a sufficiently large
line-of-sight distance (formally, the projection of the real-space and
redshift-space correlation functions are identical if the projection
is extended to infinity). Due to the small sample size of DEEP2, 3
square degrees, the statistics are not good enough to estimate
$\xi(r_p,\pi)$ out to large values of $\pi$, thus the
\cite{coil_etal:08} measurements are integrated over the range $\pm
20$ \hmpc. This $\pi_{\rm max}$ is large enough to eliminate
non-linear redshift-space effects, but large-scale infall will
contribute to the measured amplitude of $\wp$ at $r_p\gtrsim 5$ \hmpc.

Our goal in this paper is to probe the galaxy and halo mass scale at
which the transition to the red sequence begins. Thus we focus on the
DEEP2 results for the magnitude threshold sample of $M_B<-19.5$, the
faintest sample published. For the blue galaxies, the median redshift
is 0.81, while for the red galaxies the median redshift is 0.77. In
practice we use $z=0.79$ as the redshift of the combined red-blue
analytic model, noting that our results depend little on the assumed
redshift in the range specified.  As a consistency check on our
results, we also consider a bright sample of galaxies with
$M_B<-20.5$. The median redshifts for the blue and red subsamples are
0.99 and 0.88, respectively.

\cite{coil_etal:08} estimated errors from the sample-to-sample
variance from 10 independent fields.  Since the volume of this sample
is $\sim(100$\hmpc$)^3$, sample variance is a concern.  As with the
UDS sample, we also include the uncertainties in the number densities
for each sample. We estimate these uncertainties from mocks created
from the sample N-body simulation as above, but now each mock is a
cube with the same volume as the observational sample. The mocks are
created from using the best-fit HOD parameters under the assumption of
no errors in the number density. The uncertainties in the abundances
are 11\% and 13\% for the blue and red subsamples, respectively.

All the samples in \cite{coil_etal:08} are separated into blue and red
subsamples by a tilted cut in color-magnitude space,
$(U-B)=-0.023(M_B+21.62)+1.035$. Using this cut, $\sim 27\%$ of the
galaxies in the faint sample are red, while $\sim 36\%$ of the objects
in the bright samples are classified as red. As opposed to the UVJ
method of defining samples in the UDS, a single color cut can be
subject to dust contamination. We will discuss the possible
systematics of this effect in \S 6.

\begin{figure*}
\epsscale{1.2} 
\plotone{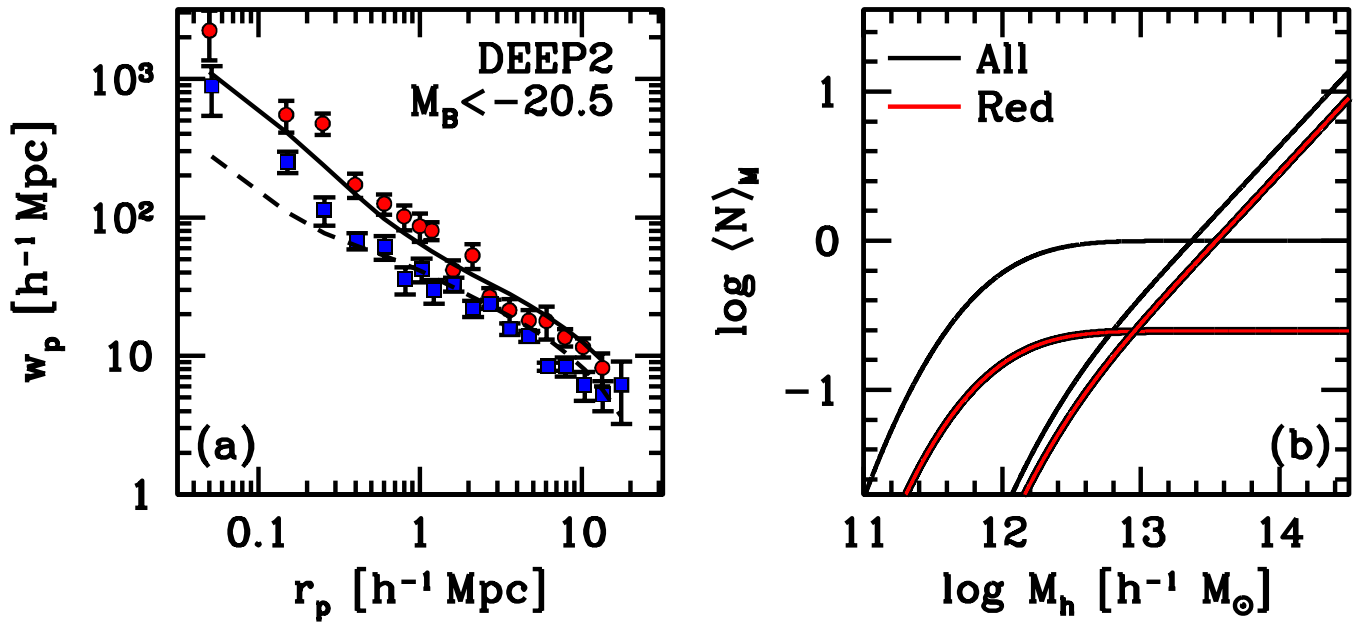}
\vspace{-10.0cm}
\caption{ \label{wp_deep_20.5} Panel(a): Measurements of the projected
  correlation function for galaxies with $M_B<-20.5$ in DEEP2 by
  \cite{coil_etal:08}. The sample has a median redshift of $z\sim
  0.9$. The circles and squares represent the measurements for red and
  blue galaxies, respectively. The solid and dashed curve represent
  the HOD model fits for red and blue galaxies, respectively. Panel
  (b): The best-fit occupation functions for all galaxies in the
  sample (thin solid curves) and red galaxies (thick curves). The
  model results imply that there is no threshold mass scale at which
  central galaxies preferentially become red; the higher clustering
  amplitude of red galaxies is produced by the high fraction of red
  satellite galaxies.}
\end{figure*}

\subsection{Clustering Measurements from COMBO-17}

The COMBO-17 survey is a photometric survey covering 0.78 deg$^2$ with
17 broad- and medium-band filters, producing high-precision
photometric redshifts (see \citealt{wolf_etal:03, wolf_etal:04} for
details of the survey). Galaxies are identified in observed-frame
$R$-band, from which P06 created a volume-limited sample of galaxies
down to $M_B<-18$. The photometric redshifts are accurate enough to
yield measurements of $\xisp$ and $\wp$ at a redshift of $z=0.6$,
presented in P06. Unlike the DEEP2 sample, in which $\wp$ is obtained
from integrating $\xisp$ out to 20 \hmpc, P06 integrate out to 100
\hmpc\ using the direct measurements, then continue out to 200 \hmpc\
using a linear theory model set to match the amplitude of the data at
$\pi=100$ \hmpc. With this approach, redshift-space distortions are
eliminated in the estimate of $\wp$, provided that the linear theory
model is accurate. Error bars are estimated from jackknife sampling of
the COMBO-17 survey. Although P06 present measurements of the
covariance matrix of both the red and blue subsamples, the matrices
themselves are too noisy to be properly inverted. We use the diagonal
errors obtained from the jackknife method in our analysis, but it
should be noted that the matrices presented in \cite{phleps_etal:06}
show that $\wp$ becomes correlated outside of $r_p\sim 1$ \hmpc. We
assume the same uncertainties in the number densities as for the faint
DEEP2 sample.

Utilizing the wide wavelength coverage of the 17 photometric bands,
\cite{phleps_etal:06} used spectral-energy distribution (SED) fitting
to break their sample of galaxies into active and passive types. The
template spectra used in the fitting comprise a two-dimensional
age/reddening sequence to remove the effect of dust on the
classification of galaxies. A $U-V$ color cut based on the
prescription of \cite{bell_etal:04} is then used to divide the sample
into red and blue galaxies. This cut is a function of $M_V$ and
redshift.

\subsection{Comparison of Surveys}

These three surveys make up a heterogeneous collection of samples that
vary in the bands, magnitude limits, and definitions of red and
blue. It is perhaps more productive to look at the number densities
rather than the luminosities---the number density of a threshold
sample of galaxies correlates strongly with the mean halo mass scale
being probed. The halo mass function also evolves from $z=1.4$ to
$z=0.6$; the abundance of massive halos increases dramatically, while
the number of $M\lesssim 10^{11}$ halos evolves only weakly. At the
redshift of the UDS, the nonlinear halo mass scales---where the bias
is approximately unity---is $1.7\times 10^{10}$ \hmsol, while at
$z=0.6$ it is $3.1\times 10^{11}$ \hmsol. At all redshifts, the halo
mass scale probed by each sample is above this nonlinear scale,
implying that any difference in the halo masses probed by red and blue
galaxies will produce differences in the large-scale clustering.

The mean redshifts, redshift ranges, and number densities of the
galaxy samples are shown in Figure \ref{number_densities}. Ignoring
the bright DEEP2 sample for the moment, the number density in our
samples increases by roughly a factor of 3 from $z\sim 1.4$ to $z\sim
0.6$. Thus COMBO-17 is probing the clustering of lower-mass halos than
either DEEP2 or the UDS. However, the fraction of galaxies that are
red or passive is roughly constant at $\sim 25\%$.

\section{Methods}

\subsection{The Halo Occupation Distribution for a Threshold Sample of
  Galaxies}

Before describing our model for the occupation of galaxy samples
defined by color, we start by defining the halo occupation of all
galaxies, red and blue together. For galaxy samples that are complete
down to a given luminosity threshold, halo occupation is broken into
two distinct parts: galaxies that reside in the center of a dark
matter halo and satellite galaxies that are within the virial radius
of the halo but are distributed throughout the halo.

For central galaxies, we parameterize the central occupation function
as

\begin{equation}
\label{e.ncen}
\ncen = \frac{1}{2}\left[ 1+\mbox{erf}\left(\frac{\log M - \log
    \mmin}{\sigmaM} \right) \right].
\end{equation}

\noindent Equation (\ref{e.ncen}) yields a smooth transition between
halos that are too small to contain galaxies bright enough to be
included in the given sample ($M\ll \mmin$) and halos that are massive
enough such that they will always contain a galaxy at their center
bright enough to be included ($M\gg \mmin$). Formally, $\mmin$ is
defined as the halo mass at which a galaxy has a 50\% probability of
containing a central galaxy in the sample. The parameter $\slogm$
controls how rapid the transition is between zero and one central
galaxies. Physically, $\slogm$ represents the scatter in halo mass at
the luminosity threshold defined by the sample, under the assumption
that this scatter takes the form of a lognormal distribution. Equation
(\ref{e.ncen}) is the mean number of central galaxies; because there
can only be one or zero central galaxies in a halo, the scatter about
that mean is defined by a nearest-integer distribution. Low-redshift
estimates of the mass-luminosity scatter yield values between 0.2 and
0.6, depending on luminosity (\citealt{tinker_etal:06_voids,
  tinker_etal:07_pvd, zheng_etal:07, vdb_etal:07, more_etal:09}). For
the UDS and COMBO-17 samples we set $\slogm=0.3$ while for both DEEP2
samples we set $\slogm=0.6$. The motivation for the higher $\slogm$
for DEEP2 springs from the observed clustering amplitude of DEEP2
galaxies; the large-scale amplitude of $\wp$ is somewhat lower than
predictions using the WMAP5 cosmology and a reasonable bias model (see
\citealt{wetzel_white:09}). This could imply larger magnitude errors
or sample variance in the data. Allowing $\slogm$ to be a free
parameter drives the best-fit values to physically unrealistic regions
on parameter space. Fixing these values, rather than allowing them to
be free parameters, does not bias our results.

For satellite galaxies, we parameterize the mean satellite occupation
function as

\begin{equation}
\label{e.nsat}
\nsat = \frac{1}{2}\left[ 1+\mbox{erf}\left(\frac{\log M - \log
    2\mmin}{\sigmaM} \right) \right]\,\left(\frac{M}{\msat}\right)^{\asat}.
\end{equation}

\noindent At halo masses well above the minimum mass scale, the number
of satellite galaxies scales as a power-law with index $\asat$ and
normalization $\msat$. This is well-motivated from the results of
hydrodynamical cosmological simulations (e.g., \citealt{white_etal:01,
  berlind_etal:03, zheng_etal:05}) as well a high-resolution
collisionless simulations (e.g., \citealt{kravtsov_etal:04,
  conroy_etal:06, wetzel_white:09}).  In equation (\ref{e.nsat}), the
satellite occupation function has a cutoff of the same functional form
as $\ncen$, but with a transition mass a factor of two larger that for
central galaxies. This cutoff prevents halos with a low probability of
containing a central galaxy from having a higher probability of
containing a satellite galaxy in the sample, and is also motivated
from the numerical results listed above. We assume that the scatter
about the mean $\nsat$ is Poisson (\citealt{kravtsov_etal:04,
  zheng_etal:05}).

We set $\asat=1$ for all samples. There is strong theoretical and
observational evidence for such a prior. High resolution dark matter
simulations find that the subhalo mass function is nearly self-similar
(\citealt{gao_etal:04, de_lucia_etal:04, kravtsov_etal:04}), thus if
satellite luminosity is tightly correlated with subhalo
mass\footnote{As noted in \cite{nagai_kravtsov:05},
  \cite{vale_ostriker:06}, \cite{conroy_etal:06}, \cite{wang_etal:06},
  and \cite{wetzel_white:09}, satellite luminosity or mass correlates
  best with the mass (or circular velocity) of the subhalo at the time
  it is accreted, rather than the time it is observed.}  the number of
satellites above a given luminosity should scale linearly. Both
cosmological hydrodynamic simulations and semi-analytic models of
galaxy formation find that this scaling is not broken with the
inclusion of baryonic physics (\citealt{white_etal:01, zheng_etal:05,
  croton_etal:06a, harker_etal:07}). Observationally, there is some
discrepancy in the measured in clusters. \cite{lin_etal:04} fin a
value of $\asat<1$, while \cite{kochanek_etal:03} find $\asat=1.10\pm
0.09$ (for a sample that contains significant overlap with the
\citealt{lin_etal:04} sample). \cite{collister_lahav:05}, using a much
larger sample of galaxy groups from the 2PIGG catalog
(\citealt{eke_etal:04}), find $\asat=0.99$. \cite{yang_etal:08} find
$\asat\approx 1$ from a group catalog constructed from the
spectroscopic galaxy sample in DR4 of the SDSS.  In the largest sample
of galaxy clusters to date, the maxBCG cluster catalog
(\citealt{koester_etal:07}), the results are also consistent with
$\asat=1$ (\citealt{hansen_etal:09}). We will discuss the effect of
leaving $\asat$ as a free parameter for both red and blue samples in
\S 6.

\begin{figure*}
\epsscale{1.2} 
\plotone{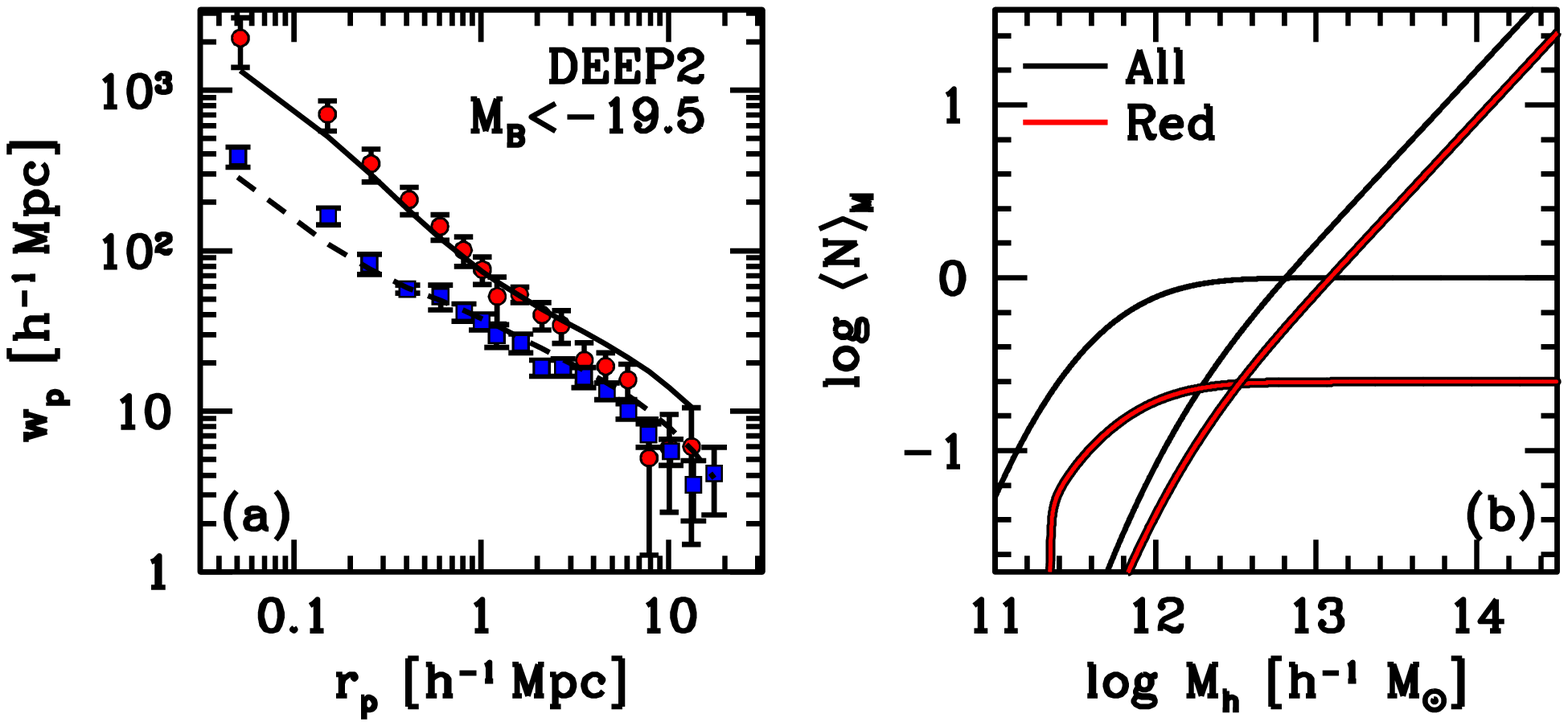}
\vspace{-10.0cm}
\caption{ \label{wp_deep_19.5} Panel (a): Measurements of the
  projected correlation function for galaxies with $M_B<-19.5$ in
  DEEP2 by \cite{coil_etal:08}. The sample has a median redshift of
  $z\sim 0.8$. The circles and squares represent the measurements for
  red and blue galaxies, respectively. The solid and dashed curve
  represent the HOD model fits for red and blue galaxies,
  respectively. Panel (b): The best-fit occupation functions for all
  galaxies in the sample (thin solid curves) and red galaxies (thick
  curves). The asymptotic fraction of central galaxies that are red,
  $\frmax$, is set to be 0.25 to match the results from Figure
  \ref{wp_deep_20.5}. As with the bright sample from DEEP2, the HOD
  results for this sample imply that there is no preferred mass scale
  for red central galaxies; increased clustering of red galaxies is
  due mostly to $\frsat$, which is nearly twice as high as the overall
  fraction of red galaxies in this sample. }
\end{figure*}

\subsection{Separating Galaxies into Red and Blue}

Many previous studies have presented halo occupation modeling of red
and blue galaxy clustering (\citealt{scranton:03,
  magliocchetti_porciani:03, zehavi_etal:05}; P06;
\citealt{tinker_etal:08_voids, skibba_sheth:09}). We will present a
somewhat modified approach to color-dependent HOD modeling that is an
extension of \cite{tinker_etal:09_drg}.

For satellite galaxies, we utilize a simple model to separate red and
blue subsamples: a constant fraction of satellites, $\frsat$, are red,
independent of halo mass. Previous models for low-redshift
color-dependent data, ie \cite{zehavi_etal:05}, allow this red
satellite fraction to vary with halo mass. But in most samples
investigated by \cite{zehavi_etal:05} and \cite{tinker_etal:08_voids},
as well as our own tests, the best-fit models have little to no
variation of $\frsat$ with $M$. In the DEEP2 galaxy group catalog of
\cite{gerke_etal:07}, the red fraction is independent of group
richness, supporting our assumption that the $\frsat$ is independent
of halo mass. We will discuss the possible biases of this assumption
in \S 6.

For central galaxies, we implement a model with considerably more
flexibility. Because the the efficiency of star formation quenching
for central galaxies may be halo mass dependent, the minimum mass
scale for red central galaxies may be much higher than that for blue
galaxies in a luminosity-threshold sample. Current models predict that
this transition could be sharp (e.g., \citealt{croton_etal:06a,
  dekel_birnboim:06, cattaneo_etal:06}), or that it could be quite
broad (e.g., \citealt{bower_etal:06, hopkins_etal:08b,
  somerville_etal:08}). At high redshift, massive halos may be more
efficient at forming stars than in the present universe, thus a high
fraction of high-mass halos may be blue (\citealt{dekel_etal:08}). We
parameterize all these possibilities with a red central fraction of
the form

\begin{equation}
\label{e.frcen}
\frcen(M) = \frmax \exp\left[\frac{-\beta\kappa\mmin}{M-\beta\mmin}\right]
\end{equation}

\noindent when $M>\beta\mmin$, and $\frcen=0$ at lower masses. The
parameter $\kappa$ governs how sharp the transition is between halos
hosting no central red galaxies and halos having an asymptotic
probability of $\frmax$ of having a red central galaxy. The parameter
$\beta$ sets the overall shift in the mass scale for red central
galaxies by setting the mass scale below which no halos host red
central galaxies.

Figure \ref{ncen_basic} shows several examples of the red occupation
function, defined as $\frcen\times\ncen$. The thick solid line shows
an example of $\ncen$, the central occupation function for all
galaxies, for the UDS sample. $\mmin$ for this sample is roughly
$10^{11.6}$ \hmsol. The thin solid curve shows a model in which
$(\frmax, \kappa, \beta) = (0.2, 0, 0.1)$, which yields a constant
$\frcen=0.2$ at all halo masses. A quantity of interest is the ratio
of the average halo mass of a red central galaxy to the average halo
mass of a blue central galaxy, defined as

\begin{equation}
\label{e.mred_mblue}
\frac{M_{\rm red}}{M_{\rm blue}} = \frac{n_{\rm Rcen}^{-1}\int dM M n(M) \frcen \times\ncen}{n_{\rm Bcen}^{-1}\int dM M n(M) (1-\frcen) \times\ncen}
\end{equation}

\noindent where $n(M)$ is the halo mass function, and $n_{\rm Rcen}$
and $n_{\rm Bcen}$ are the number density of red and blue central
galaxies, respectively, that are calculated from the HOD by, eg, 

\begin{equation}
n_{\rm Rcen} = \int dM n(M) \frcen \times\ncen.
\end{equation}

\noindent For the thin solid curve, $\mredmblue=1$ by definition. Red
central galaxies are essentially a random subset of all central
galaxies. The dotted curve shows a model in which $(\frmax, \kappa,
\beta) = (0.75, 24, 0.75)$. This model produces a very broad
transition between halos having a negligible probability of containing
a red central galaxy to central galaxies being majority red. The
number density of red centrals is the same as the solid curve, but now
the mass ratio between red and blue central galaxies is 3.1. This
function would represent a merger-induced red central model. The
dashed curve shows a model in which $(\frmax, \kappa, \beta) = (0.8,
0.057, 2.5)$. In this model, the transition between blue and red
central galaxies is nearly instantaneous at $M=2.5\mmin$. Once again,
$n_{\rm Rcen}$ is the same as in the previous two models, but
$\mredmblue$ is increased to 4.6. A function of this form represents
the critical mass scale scenario.

\begin{figure*}
\epsscale{1.2} 
\plotone{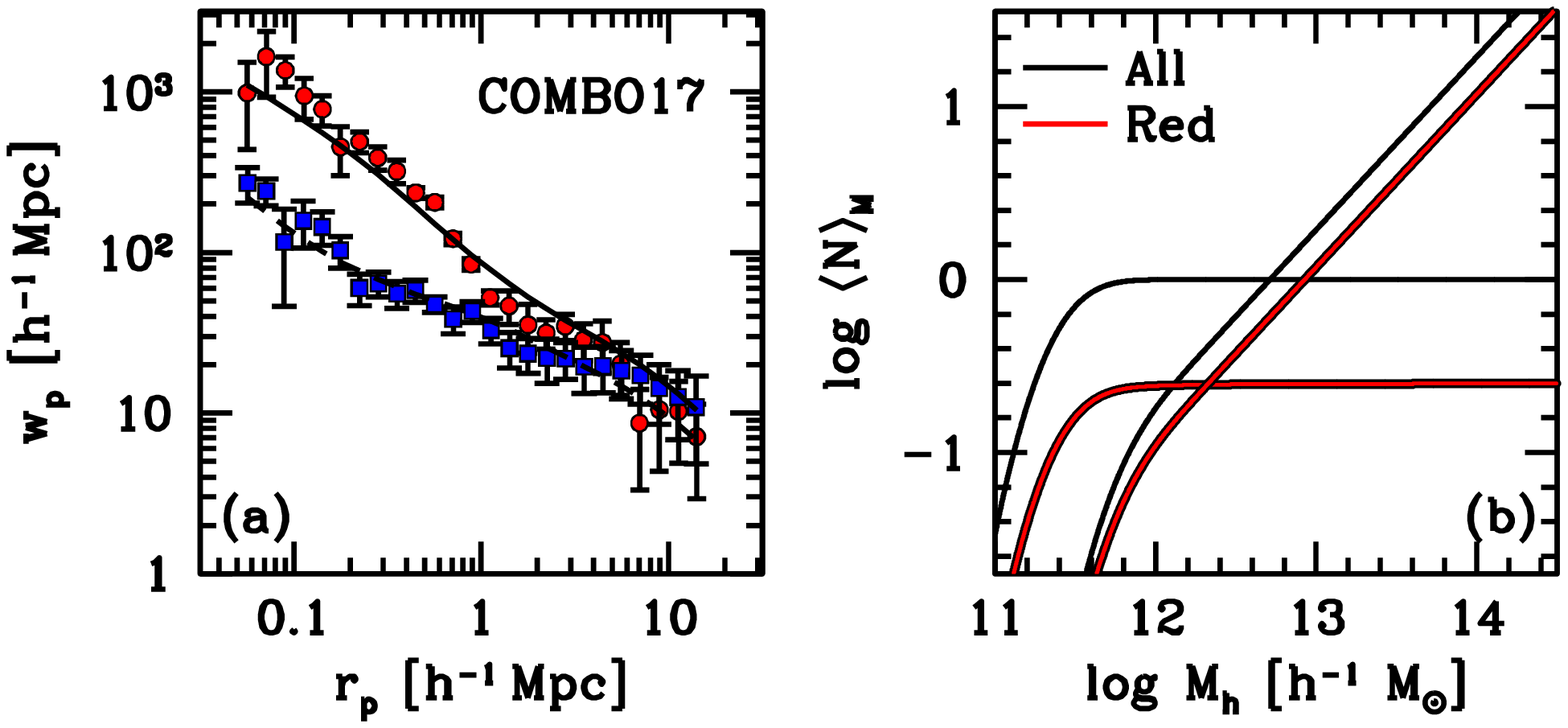}
\vspace{-10.0cm}
\caption{ \label{wp_combo17} Panel (a): Measurements of the projected
  correlation function for galaxies in COMBO-17 by
  \cite{phleps_etal:06}. The median redshift of the sample is $z\sim
  0.6$. The circles and squares represent the measurements for red and
  blue galaxies, respectively. The solid and dashed curve represent
  the HOD model fits for red and blue galaxies, respectively. Panel
  (b): The best-fit occupation functions for all galaxies in the
  sample (thin solid curves) and red galaxies (thick curves). The halo
  mass ratio of red central galaxies, $\mredmblue$, is 1.27, in
  agreement with the results from both DEEP2 samples.}
\end{figure*}

\subsection{Calculating Observables with the Model}

To calculate the galaxy autocorrelation function, $\xi(r)$, from a
given HOD, we use the analytic model described in the Appendix of
\cite{tinker_etal:05} (see also \citealt{zheng:04}). As a brief
description, the correlation function is broken into two parts: a
one-halo term in which pairs of galaxies reside within a single halo,
and a two-halo term in which pairs come from two distinct halos. The
one-halo term dominates the correlation function at separations
$r\lesssim 1$ \hmpc, while the two-halo term contributes nearly all
pairs at $r\gtrsim 1$ \hmpc. The shape of the one-halo term is
influenced by the radial distribution of satellite galaxies, which we
assume to be the same as the dark matter. Specifically, we use the
halo concentration-mass relation from \cite{zhao_etal:08}. More
importantly, the one-halo term is sensitive to the overall fraction of
galaxies that are satellites (see the discussions in
\citealt{zheng_etal:08} and \citealt{tinker_etal:09_drg}). The
satellite fraction is influenced both by the shape of the mass
function and the form of the HOD. More satellites will increase the
large-scale bias of a sample, but because the number of
satellite-satellite pairs within a single halo increase as $\nsat^2$,
the effect of the small-scale clustering is more dramatic. At large
scales, the shape of the two halo term is the same as the linear
matter correlation function. At intermediate scales, $r\lesssim 3$
\hmpc, halo bias deviates from a simple linear approximation, and a
scale-dependent term is required (\citealt{tinker_etal:09_bias}). When
implementing the analytic model for $\xi(r)$, we use the halo mass
function of \cite{tinker_etal:08_mf} and the halo bias function of
\cite{tinker_etal:09_bias}. We assume that a halo is defined to be a
spherical object with a mean interior density 200 times the background
density.

For the UDS, the angular correlation function, $\wth$, is
measured. This quantity is defined as

\begin{equation}
\label{e.wth}
\wth = \int dz\,N^2(z)\, \frac{dr}{dz} \int dx\, \xi\left(\sqrt{x^2 + r^2\theta^2}\right),
\end{equation}

\noindent where $N(z)$ is the normalized redshift distribution of the
galaxy sample, $r$ is the comoving radial distance at redshift $z$ and
$dr/dz = (c/H_0)/\sqrt{\om(1+z)^3+\oml}$. As stated in \S 2.1, we use
the estimate of $N(z)$ from \cite{williams_etal:09} that incorporates
photometric redshift errors.

For DEEP2, we calculate the projected correlation function,

\begin{equation}
\label{e.wp}
\wp = 2\int_0^{\pi_{\rm max}}\xi(r_p,\pi)d\pi.
\end{equation}

\noindent As mentioned in \S 2.2 we set $\pi_{\rm max}=20$ \hmpc, the
same as in the measurements. For the one-halo term, 20 \hmpc\ is
sufficient to eliminate redshift-space effects, therefore we use the
isotropic real-space one-halo term in equation (\ref{e.wp}). For the
two-halo term, coherent infall is still a concern.  We take this effect
into account by using the linear theory model of \cite{kaiser:87} when
calculating the two-halo term. Although linear theory is not a fully
descriptive model of $\xisp$ at separations of a few \hmpc, for a
projected quantity like $\wp$ we find that it compares well to galaxy
mocks constructed from N-body simulations in which $\wp$ is estimated
with $\pi_{\rm max}=20$ enforced. See \cite{hamilton:98} and
\cite{hawkins_etal:03} for a thorough discussion and application of
the \cite{kaiser:87} model to an unparameterized correlation function.

Because the COMBO-17 data integrate to $\pi_{\rm max}=200$ \hmpc, the
effect of redshift-space distortion is negligible. To calculate $\wp$
for the COMBO-17 data, no accounting of redshift-space effects are
required and we use the isotropic, real-space $\xi(r)$ in equation
\ref{e.wp}, and integrate along the line of site until convergence.

For each sample, we use the Monte Carlo Markov Chain (MCMC) method to
probe the likelihood distribution of the full parameter space. We have
six free parameters. The first two, $\mmin$ and $\msat$, specify
$\navg$ for the full sample of galaxies. We have one parameter,
$\frsat$, to split the satellite galaxies into red and blue
subsamples. The last three parameters, $\beta$, $\kappa$, and $\frmax$,
determine the fraction of central galaxies that are red as a function
of halo mass. These parameters specify not just $\xi(r)$ but also the
abundance of blue and red galaxies. We minimize the total $\chi^2$ of
the model, which we obtain from the sum of the $\chi^2$ for the red
and blue clustering measurements and the $\chi^2$ for the galaxy
abundances. The best-fit parameters for each sample are given in Table
1.

\begin{figure*}
\epsscale{1.0} 
\plotone{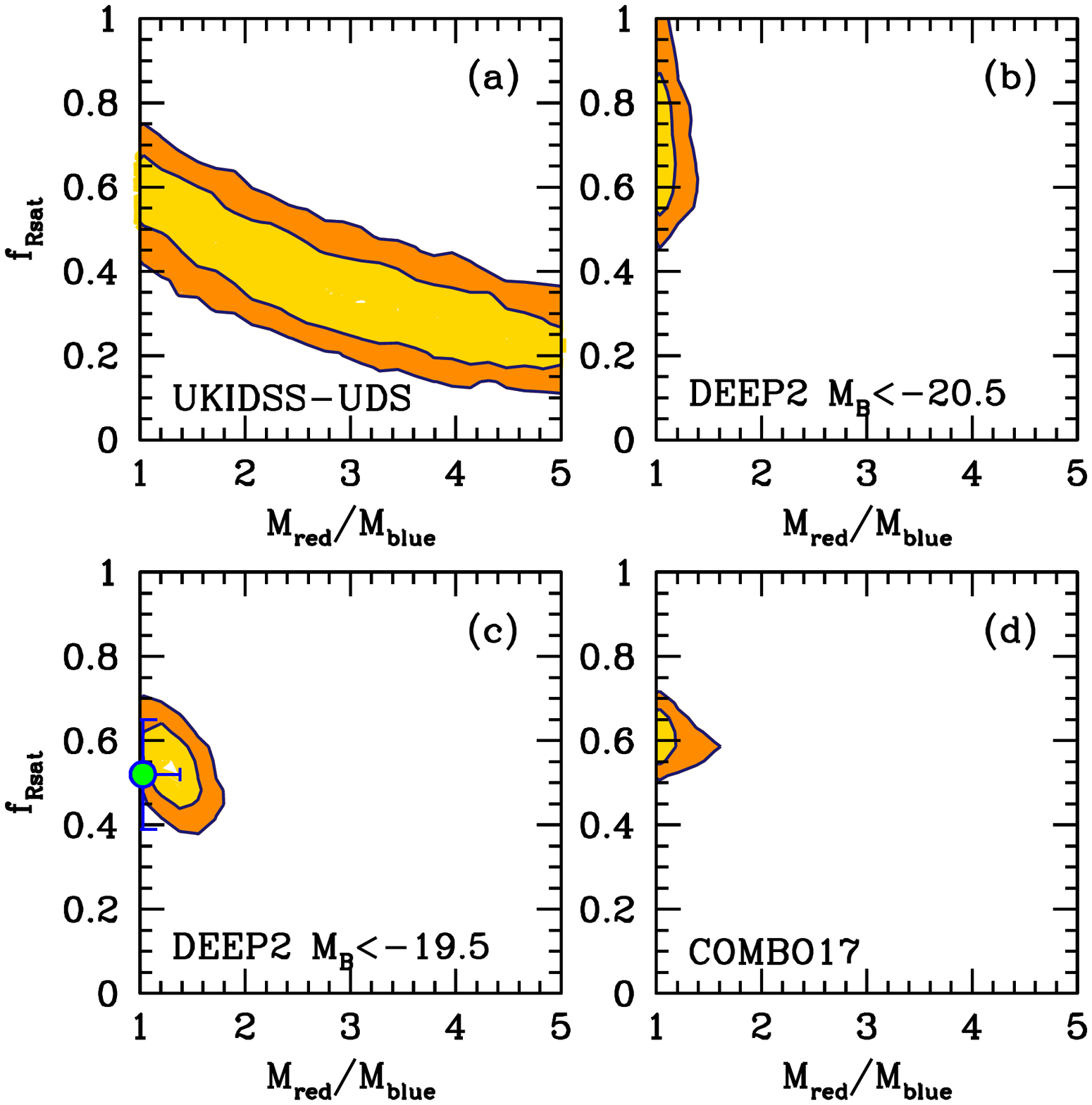}
\caption{ \label{mcen_fred} Constraints in the $\mredmblue$-$\frsat$
  plane for the four samples analyzed in this paper. Contours
  represent 1- and $2-\sigma$ constraints. In panel (a), showing the
  constraints from the UDS clustering data, the degeneracy between the
  central mass ratio and the fraction of red satellites is clear. For
  the other three samples, $\mredmblue$ is consistent with unity, with
  a red satellites fraction of $\sim 0.6$. In Panel (c), showing the
  constraints for the faint DEEP2 sample, the circle with errorbars
  represents the results of a model in which $\frmax$ is allowed to be
  a free parameter. In the contours, $\frmax=0.25$ to agree with the
  results from the bright DEEP2 sample. Errorbars are $2\sigma$.}
\end{figure*}

\begin{figure}
\epsscale{1.0}
\plotone{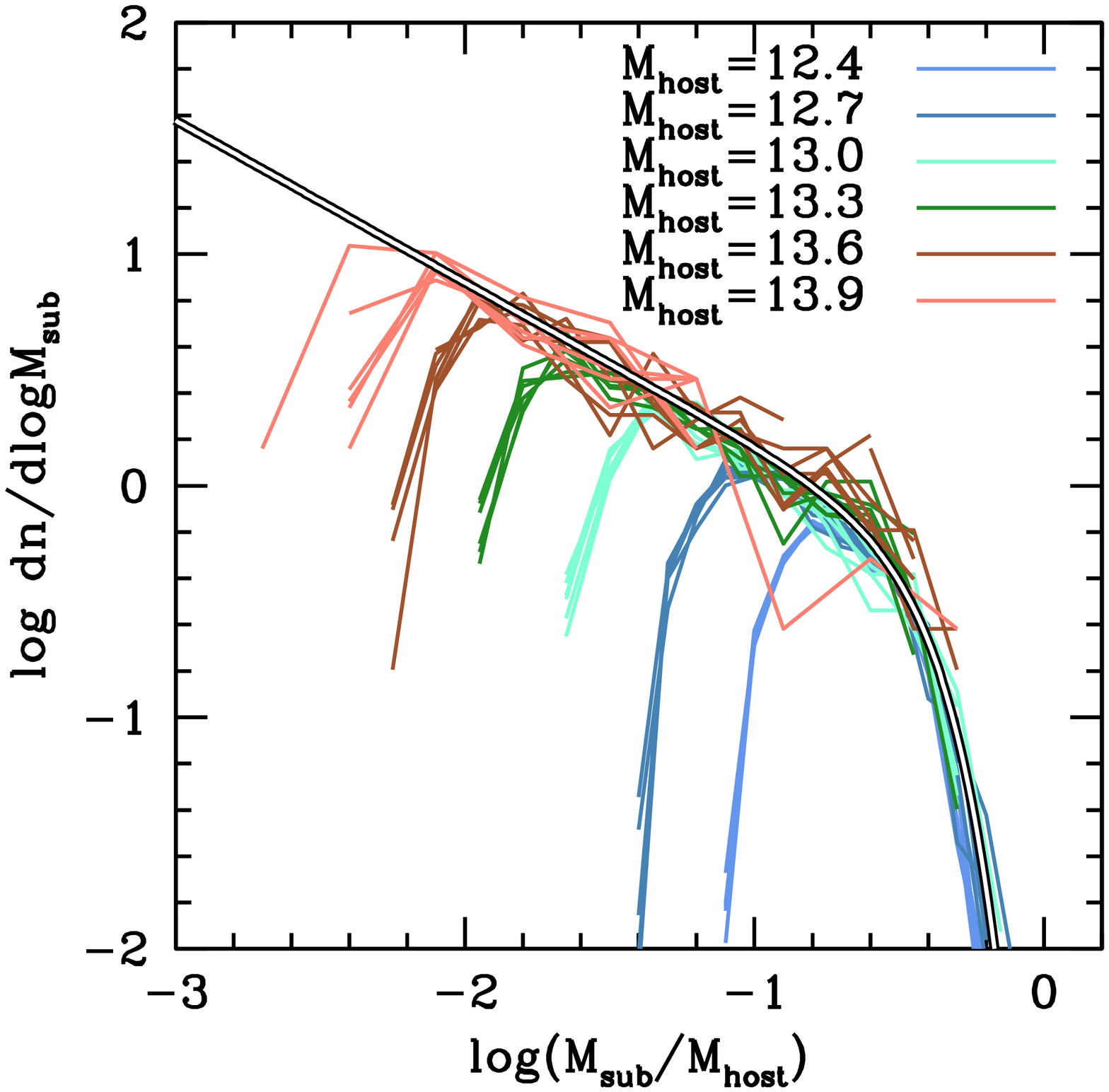}
\caption{ \label{smf} The subhalo mass function from a numerical
  simulation presented in \cite{wetzel_white:09}. Here $\msub$ is the
  mass of the subhalo at the time of accretion. The thin curves
  represent the N-body results, while the thick curve is the fitting
  function given in Equation (\ref{e.smf}). For the numerical results,
  each color represents a different host halo mass bin, ranging from
  $\log\mhost=12.4$ to $\log\mhost=13.9$. For each bin in $\mhost$,
  the six lines show results for six redshift outputs ranging from
  $z=0.8$ to $z=1.6$. The turnover at low $\msub/\mhost$ for each mass
  bin is due to the fixed particle limit in the subhalo catalog; we
  use only subhalos with more than 2000 particles at infall to avoid
  artificial numerical disruption. The overlapping results for each
  mass bin show that the subhalo mass function is universal over a
  wide range of halo mass and redshift. }
\end{figure}

\section{Results}

\subsection{Clustering in the UDS}

Figure \ref{wtheta_uds}a shows the best-fit model against the $\wth$
data, and Figure \ref{wtheta_uds}b shows the HOD for all galaxies and
the quiescent subsample. The relative bias between the star-forming
and quiescent galaxies is high enough that the best-fitting model is
on the extreme edge of allowed parameter space; to increase the
clustering of quiescent galaxies relative to the star-forming
subsample, the best-fit $\frcen$ rapidly rises from 0 to 1 at a
``critical'' mass threshold of $M=10^{12.2}$ \hmsol, yielding a halo
mass ratio of red to blue centrals of $\mredmblue\sim 6$.  Because of
the high central mass ratio, the clustering amplitude of red galaxies
is boosted significantly. This fact, along with the high fraction of
centrals that are red, forces the fraction of red satellite galaxies
down to $\frsat \sim 22\%$, which is lower than the overall red
fraction of galaxies.

However, there is a strong degeneracy axis between $\mredmblue$ and
$\frsat$. Because the number of satellites increases linearly with
halo mass, increasing $\frsat$ also increases the large-scale bias of
red galaxies without any increase the halo mass scale of red central
galaxies. This, combined with the large error bars on the data, make
the constraints on the parameters of the HOD model weak. Within the
$2\sigma$ confidence limit, a model in which $\frsat=0.59$ and
$\mredmblue=1.01$ is acceptable. This model is shown with the dashed
line in Figure \ref{wtheta_uds}b. We will discuss this degeneracy
and its implications in more detail \S 5.

\subsection{Bright galaxies in DEEP2}

Figure \ref{wp_deep_20.5} presents results for DEEP2 galaxies in the
$M_B<-20.5$ sample. Figure \ref{wp_deep_20.5}a shows both the measured
clustering signal and the best-fit HOD model. Figure
\ref{wp_deep_20.5}b shows the HOD for both the full galaxy sample and
the red subsample. At all scales, the clustering of red galaxies is
higher relative to blue galaxies. However, the $w_p({\rm
  red})/w_p({\rm blue})$ ratio is significantly smaller than that
found in lower redshift samples such as SDSS (\citealt{zehavi_etal:05,
  li_etal:06}) and 2dFGRS (\citealt{norberg_etal:02,
  madgwick_etal:03}). In fact, the the increased clustering of red
galaxies in the two-halo term is best accounted for by a model in
which red central galaxies have no dependence on halo mass
($\mredmblue=1.04$), but with a high fraction of satellite galaxies
being red ($\frsat=0.66$), twice the global fraction of red galaxies
in this sample. Unlike the analysis of the UDS clustering, the
constraints on the model parameters are strong.

The $\chi^2$ value, at 64.4 for 29 degrees of freedom, is not ideal
but we note that without a proper covariance matrix the $\chi^2$
values are not fully robust. The $\chi^2$ for the blue galaxies (39.1)
is significantly larger than for the red (18.3). Visually, it appears
that this is a result of the poor fit to the blue $\wp$ in the
one-halo term, but point-to-point scatter is also a concern; two
points, the second and fourteenth, contribute a $\chi^2$ of 22. To
improve the fit at small scales, the amplitude of the one-halo term
for both red galaxies and blue galaxies would need to increase. This
is obtained by increasing the number of satellite galaxies, but it
would come at the cost of increasing the amplitude of the large-scale
bias as well. The best-fit model is somewhat above the data at large
scales ($r_p\gtrsim 2$ \hmpc), and increasing this disparity would
decrease the quality of the fit (see also
\citealt{wetzel_white:09}). \cite{zheng_etal:07}, also using HOD
modeling, produced fits that accurately reproduced the large-scale
clustering of DEEP2 galaxies, but this result is due to the choice of
cosmology: the WMAP1-type cosmology utilized in \cite{zheng_etal:07}
has significantly less large scale power due its lack of tilt and
higher matter density. We find that we are able to reproduce
the \cite{zheng_etal:07} result when implementing the WMAP1
cosmology. 

In this paper we are most interested in probing the halo mass scale at
which star formation for red central galaxies becomes quenched, if
there is one. Analysis of bright galaxies probes a higher halo mass
scale, but this sample is useful for a number of reasons. First, it is
a consistency check on the results obtained from the faint
galaxies. Second, because the samples in \cite{coil_etal:08} are
threshold samples, we know that the $\frcen$ for the {\it faint}
sample can never be lower than $\frcen$ for the {\it bright} sample at
a given halo mass. It is also highly unlikely that $\frcen$ for the
fainter sample would be higher than that of the brighter sample at
$M\gtrsim 10^{14}$ \hmsol\ unless magnitude errors were extreme. The
bright sample is fully contained within the faint sample; the central
galaxies in high-mass halos are the same galaxies that produce the
clustering in the bright sample. Thus we fix the value of $\frmax$ in
the faint sample to be the the best-fit value of from the $M_B<-20.5$
sample, 0.25, ensuring that $\frcen$ for the faint sample will be the
same as that in the bright sample in the overlapping halo mass range.

\begin{figure}
\epsscale{1.0}
\plotone{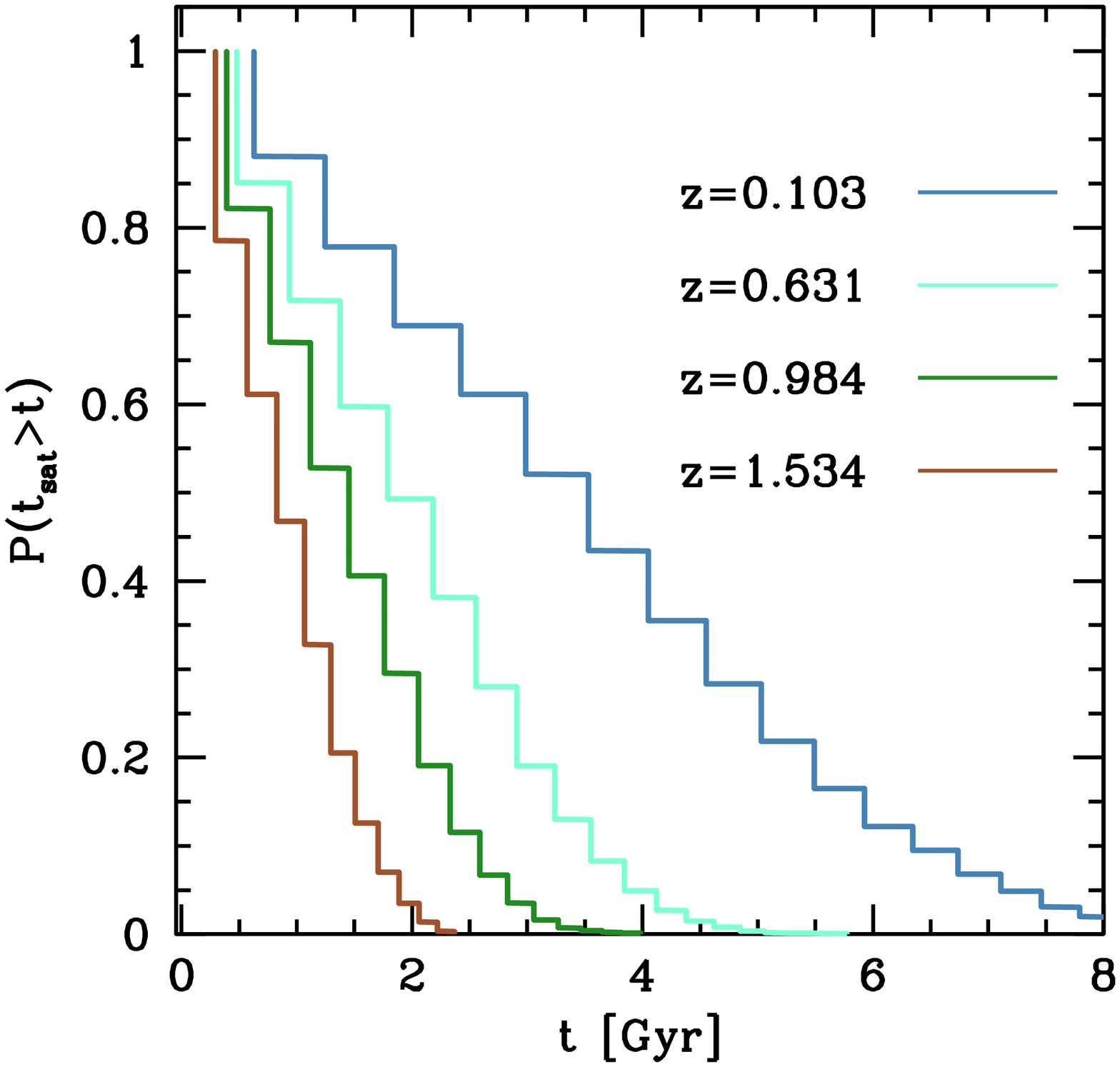}
\caption{ \label{tsat} The distribution of subhalo ages, defined as
  the time since a subhalo was accreted onto its parent halo. Results
  are from the simulations of \cite{wetzel_white:09}. The $y$-axis
  indicates the fraction of subhalos with accretion times longer than
  the value $t$. As redshift increases, the mean age of a subhalo
  becomes younger because dynamical friction is more efficient at
  higher redshifts due to the higher density of most halos. The halo
  catalogs span the redshift range $z=0$ to $z=1.6$, but in this plot
  we show only four output redshifts to avoid crowding.  }
\end{figure}

\subsection{Faint galaxies in DEEP2}

Figure \ref{wp_deep_19.5}a presents the clustering of $M_B<-19.5$ in
DEEP2, along with the model with the lowest $\chi^2$ from the MCMC
chain. As with the bright galaxies, the clustering of red galaxies in
this sample is stronger than that of the blue galaxies at all scales
(the last three data points in the red $\wp$ that have low amplitudes
are most likely a result of sample variance), but the relative bias
between red and blue samples at large scales is only moderately above
unity (\citealt{coil_etal:08} calculate a relative bias of
1.25). Thus, the best-fitting HOD in Figure \ref{wp_deep_19.5}b is
similar to that of the bright galaxies; $\frcen$ is nearly independent
of halo mass, yielding $\mredmblue=1.48$. The central occupation
function for red galaxies contains a sharp cutoff at $M=10^{11.4}$
\hmsol, which is the feature that causes the value of $\mredmblue$ to
be larger than one. This cutoff is required by the constraints on the
number density rather than on the clustering; without such a cutoff,
the space density of red galaxies would be too high. The fraction of
satellites that are red is 52\%, twice the overall red fraction of
this sample.

Visually, the fit for this sample appears to match the data better
than with the brighter sample, but the $\chi^2$ is roughly the same,
at 63.7 (although for one less degree of freedom because $\frmax$ has
now been fixed to 0.25). The larger number density of this sample
yields smaller error bars in the 1-halo term, and once again
point-to-point variance is larger than the errors for several data
points: two data points account for 27\% of the total $\chi^2$. When
using the WMAP1 cosmology to model the red and blue, the $\chi^2$ of
the fit is 28.9 (yielding $\chi^2/\nu=0.93$), a significant
improvement on the results discussed above. However, the constraints
on $\frsat$ and $\mredmblue$ are unchanged. An additional difference
between the \cite{zheng_etal:07} analysis and that of this paper is
the freedom in $\asat$. We will discuss this is more detail in \S
6. 

As a check on these results, we have performed the HOD analysis of the
faint sample without the prior on $\frmax$ obtained from the analysis
of the bright sample. The results are consistent within their
1-$\sigma$ contours, which we will show in Figure \ref{mcen_fred}. In
this test, $\frsat$ was nearly unchanged, while the best-fit
$\mredmblue$ {\it decreased} to be 1.0, the same as for the bright
galaxies. The best-fit value of $\frmax$ is 0.16, rather than 0.25,
which removes the need for the sharp cutoff in $\frcen$ at $10^{11.4}$
\hmsol\ in order to match the number density.

\subsection{Clustering Analysis for COMBO-17}

Figure \ref{wp_combo17}a and \ref{wp_combo17}b show the clustering
results from COMBO-17 and the resulting halo occupation functions,
respectively. The $\wp$ measurements for red and blue galaxies are
similar to those in DEEP2; at small scales, the clustering of red
galaxies is strongly enhanced relative to blue galaxies, but in the
two-halo term the difference in the clustering is minimal, with sample
variance muddling the comparison at $r_p\gtrsim 10$ \hmpc. This
results in a best-fit HOD that is the same as those found in the DEEP2
sample; $\mredmblue=1.01$ and $\frsat=0.58$, which is more than twice
as high as the overall red fraction of the sample.

The reduced $\chi^2$ value of the best fit model is 1.9. The $\chi^2$
for the blue $\wp$ is 14.7 for 25 data points. The source of the high
$\chi^2$ is localized to the transition region between 1-halo and
2-halo clustering in the red $\wp$; half of the total $\chi^2$ value
is accrued between $0.5<r_p<1.1$, where the clustering amplitude drops
rapidly. P06 have also performed HOD analysis of these data, obtaining
a much better $\chi^2$ value than that found here, 35.9 versus
88.3. There are several reasons for the lower $\chi^2$ value from the
P06 analysis. First, P06 have a different cosmological model with no
tilt, which reduces the large-scale power. Second, P06 leave $\s8$ as
a free parameter, increasing the freedom of their cosmological
model. This parameter is left free for {\it both} the red and blue
subsamples, yielding best-fit values of $\s8$ of $0.84\pm 0.08$ and
$1.19 \pm 0.09$ for the red and blue subsamples, respectively. The
lower $\s8$ value for the red model fit is required by the low
relative bias of blue and red galaxies in the data. Third, and most
important, P06 implement a physically unrealistic HOD model. P06
assume the same form of the HOD for both blue and red subsamples: a
power-law in which $\navg=(M/M_0)^\beta$ for $M>M_0$. The power-law
index $\beta$ is left as a free parameter. This model is unrealistic
in that it does not reproduce known aspects of halo occupation
parameterized in equations (\ref{e.ncen}) and (\ref{e.nsat}): a
self-consistent separation of central and satellite galaxies; a
``shoulder'' in $\navg$ in the halo mass range between $\mmin$ and
$\msat$ (nearly a factor of 20 in halo mass), in which the scatter in
$\navg$ is sub-Poisson because satellites are a minority of the
galaxies. Lastly, when modeling the red and blue galaxies separately,
P06 does not require that the number of central red galaxies and
central blue galaxies be $\le 1$ at fixed halo mass. P06 stipulate
that if $\navg>1$, the `first' galaxy is central and the remaining
objects are satellites. However, this is done for {\it both} red and
blue occupation functions. The implication is that there are {\it two}
central galaxies per halo, one red and one blue, above $M_{\rm
  0,red}$. When implementing an HOD that is similar to equations
(\ref{e.ncen}) and (\ref{e.nsat}), P06 see no difference in the
$\chi^2$ for the blue galaxies, but the $\chi^2$ for the red galaxies
increases by 20. See \cite{tinker_etal:09_drg} for a full discussion
of the implications of an HOD model of the type utilized by P06.

\subsection{Parameter Constraints for All the Samples}

Figure \ref{mcen_fred} shows the constraints in the
$\mredmblue$-$\frsat$ plane. The $x$-axis does not extend below
$\mredmblue=1$ because that is the minimum value allowed from equation
(\ref{e.frcen}). For the UDS data, the lack of significant constraints
is clear from the strong degeneracy between $\frsat$ and $\mredmblue$;
with a larger fraction of red satellite galaxies, the need for a large
ratio between the masses of red and blue central galaxies is
reduced. This is a consequence of the small volume of the
sample and the use of angular clustering.

For the next three samples, the results are markedly different. All of
the samples are consistent, within the 1-$\sigma$ constraints, with
red central galaxies being a random subsample of all central
galaxies. A value of $\mredmblue=2$ is excluded at more than
2-$\sigma$ for all the samples. The fraction of red satellite galaxies
is high, being consistent with 60\% in all samples.  Note that
$\frsat=0.6$ is consistent with the results from the UDS data, in
which case $\mredmblue \approx 1$.  Thus, if there is no strong
evolution in the red satellite fraction from $z \approx 0.9$ to $1.4$,
the UDS data also suggest that red central galaxies are a random
subsample of all galaxies.

\begin{figure}
\epsscale{1.0} 
\plotone{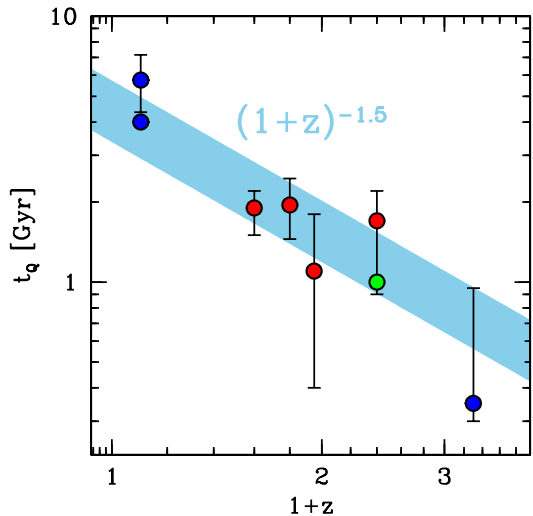}
\caption{ \label{tquench} Timescale for galaxies to transition from
  blue to red, or from star-forming to quiescent. All error bars are
  $2\sigma$. The middle four (red) data points represent, in order of
  increasing redshift, COMBO-17, DEEP2 faint, DEEP2 bright, and the
  UDS. The blue data points are taken from other papers. The datum at
  $z=2.3$ is from the \cite{tinker_etal:09_drg} analysis of the
  clustering of DRGs. The upper datum at $z=0.1$ is from
  \cite{wang_etal:07}, in which the exponential decay of star
  formation in satellites was constrained to be 2.5 Gyr. We assume
  that satellites are ``quenched'' when the star formation rate
  decreases by a factor of 10, or 2.3 e-folding times. The lower datum
  at $z=0.1$ is an estimate of the timescale for transitioning to the
  red sequence using the clustering data and analysis in
  \cite{zehavi_etal:05}. The green circle at $z=1.4$ is the quenching
  timescale inferred from the UDS data if we assume that
  $\mredmblue=1$, in agreement with the other three samples. The
  shaded band is a power-law in which the quenching timescale varies
  with the dynamical time of dark matter halos, $(1+z)^{-1.5}$. }
\end{figure}


\begin{deluxetable*}{lcccccccccc}
\tablecolumns{11} 
\tablewidth{32pc} 
\tablecaption{Parameters for Best-Fit Models} 
\tablehead{ \colhead{Sample} & \colhead{$\frsat$} & \colhead{$\frmax$} & \colhead{$\kappa$} & \colhead{$\log \msat$} & \colhead{$\log \beta$} & \colhead{$\log \mmin$} & \colhead{$\nred$} & \colhead{$\nblue$} & \colhead{$\nu$} & \colhead{$\chi^2$} }
\startdata

UDS & 0.22 & 0.98 & 2.47 & 12.803 & -1.815 & 11.838 & $1.10\times 10^{-3}$ & $3.17\times 10^{-3}$ & 19 & 10.8 \\
DEEP2 -20.5 & 0.66 & 0.25 & 0.04 & 13.365 & -0.363 & 11.882 & $8.46\times 10^{-4}$ & $2.13 \times 10^{-3}$ & 29 & 64.4 \\
DEEP2 -19.5 & 0.52 & 0.25 & 0.45 & 12.800 & -1.891 & 11.682 & $2.37\times 10^{-3}$ & $8.00 \times 10^{-3}$ & 30 & 63.7 \\
COMBO-17 & 0.59 & 0.17 & 0.13 & 12.694 & 0.050 & 11.391 & $3.96\times 10^{-3}$ & $1.23\times 10^{-2}$ & 46 & 86.1 \\

\enddata
\tablecomments{The units of $\mmin$ and $\msat$ are \hmsol. $\nu$ is the number of degrees of freedom in the sample. }
\end{deluxetable*}

\section{The Timescale for Satellite Transformation}

\subsection{A Simple Model}

Galaxies form within dark matter halos; the halo occupation formalism
is built upon this concept. High resolution collisionless N-body
simulations track dark matter halos after they have been accreted onto
larger halos. Models that associate these {\it subhalos} with the
satellite galaxies that constitute groups and clusters have been
successful at reproducing measured clustering of galaxies at various
redshifts (e.g., \citealt{kravtsov_etal:04, conroy_etal:06,
  wang_etal:06, wang_etal:07, marin_etal:08, moster_etal:09,
  wetzel_white:09}). In these models, the known unknown is the exact
relationship between halo and subhalo mass and the luminosity (or
other galaxy properties, such as stellar mass) of the galaxy. A simple
and successful approach has been to assume a monotonic relationship
between galaxy luminosity and halo mass, where the exact functional
form of the mass-to-light ratio is constrained to match the luminosity
function of galaxies. Although there is known to be some scatter in
halo mass at fixed galaxy properties (\citealt{more_etal:09}), this
monotonic approach has proved successful and useful in myriad studies,
producing results quantitatively similar to those that incorporate
scatter.

In the results in \S 4, we have made no assumptions about the
population of dark matter subhalos within distinct halos. We have also
made few assumptions about the $M/L$ ratio as a function of halo mass
because the samples we have analyzed have are defined with a threshold
luminosity, rather than being broken into multiple luminosity
bins. However, to make inferences about the evolution of satellite
galaxies we must make use of N-body simulations that keep track of
merger rates and the lifetimes of subhalos. For each observational
sample, we have determined the total number of satellite galaxies in
each halo, as well as the fraction of those galaxies that are
classified as red or quenched. To determine the quenching time of a
galaxy once it is accreted, a simple approach is to consider all
subhalos in a given halo that contain galaxies in the sample. Using
the monotonic $M/L$ ratio, this would simply be the $\nsat$ most
massive subhalos (where the subhalos are ranked by their mass {\it at
  the time of accretion}, not their present-day mass, which is
affected by orbit-dependent tidal stripping; see, e.g.,
\citealt{conroy_etal:06, moster_etal:09, wetzel_white:09}). These
subhalos are then ranked by the time elapsed between accretion and the
redshift of observation. Assuming that the oldest subhalos are the
ones that have had their star formation quenched and migrated to the
red sequence, the value of $\frsat$ sets the quenching timescale.

One complication is that some satellite galaxies may have been red
{\it before} they became satellites. The results from all the samples
in \S 3 indicate that, even at low halo masses, the fraction of field
red galaxies is 20-30\%. The mechanism by which these galaxies
transitioned to the red sequence is unrelated to ram-pressure
stripping or tidal stripping.  A model in
which field red galaxies are quenched by tidal interactions with
nearby massive structures is also disfavored; an
environmentally-dependent quenching mechanism of this type would
enhance the large-scale clustering of red galaxies. We have already
noted that the relative bias of red to blue galaxies is difficult to
match even if the probability of a central galaxy being red is
independent of environment. An environmentally-dependent $\frcen$ is
also disfavored at $z=0$ (\citealt{tinker_etal:08_voids}).

To remove the contribution of satellite galaxies that were red before
accretion, we determine the quenched fraction of satellite galaxies,
$\fq$, by

\begin{equation}
\label{e.fq}
\fq = \frac{\frsat \nsatdens - \nprev}{\nsatdens - \nprev},
\end{equation}

\noindent where $\nsatdens$ is the number density of satellites and
$\nprev$ is the number density of previously quenched satellites,
given by

\begin{equation}
\label{e.nprev}
\nprev = \int dM\,n(M)\, \int d\msub\,n_{\rm sub}(\msub|M)\frcen(\msub)\times \langle N_{\rm cen}\rangle_{M{\rm sub}}
\end{equation}

\noindent where $n_{\rm sub}(\msub|M)$ is the subhalo mass function
(where once again $\msub$ is defined as the mass at the time of
accretion), and $\frcen(\msub)$ is the central red fraction determined
for each sample. Note that $n_{\rm sub}(\msub|M)$, in our definition,
does not have units of volume but rather is the number of subhalos
within a given parent halo of mass $M$. We will describe our fitting
function for this quantity in the following subsection.  Here, we
assume that the central galaxy red fraction does not strongly evolve
since the time of satellite accretion, supported by the mild evolution
in $\frcen$ across samples of \S 3.  Equation (\ref{e.fq}) yields a
quenched fraction that is somewhat smaller than the overall red
fraction. For three of the four samples, $\frcen$ is nearly
independent of mass and $\nprev \approx \frmax\nsatdens$. For the
best-fit model obtained for the UDS, $\frcen$ depends strongly on
mass. However, in this model, the difference between $\frsat$ and
$\fq$ is minimal; in that particular model, $\frcen=0$ below $M\sim
10^{12.2}$ \hmsol. Most satellites in this sample are below this mass,
thus the fraction of subhalos that became red before accretion is
small. Recall, however, that the constraints on $\frcen$ in the UDS
are broad.

\subsection{N-body Simulations}

To implement the model in the previous subsection, we require two
quantities: the subhalo mass function and the distribution of
accretion times for subhalos. We use the halo and subhalo catalogs
from the high-resolution N-body simulation analyzed in
\cite{wetzel_white:09}. The simulation is of the same cosmology as
that assumed here. The volume of the simulation is $200$ \hmpc\ per
side, containing 1500$^3$ particles, with a particle mass of $1.6
\times 10^8\,h^{-1}M_\odot$ and a force resolution of $3\,h^{-1}$~kpc.
Halos are identified using the friends-of-friends halo finding
algorithm with a linking length of 0.168 times the mean interparticle
separation. Halos defined in this way have different masses that halos
defined by $\Delta=200$ (\citealt{white:02, tinker_etal:08_mf}), as
done in all analytic calculations in this paper. We convert from FOF
mass to $\Delta=200$ mass with a factor of 1.3.  See
\cite{wetzel_etal:09b} and \cite{wetzel_white:09} for details of
subhalo finding and tracking.

Figure \ref{smf} shows the results for the subhalo mass function at
six different redshifts between $z=0.9$ and $z=1.6$. The results are
plotted as a function of $\msub/\mhost$, in which units $dN/d\ln\msub$
is independent of host halo mass. This self-similarity of the subhalo
mass function has been shown before for subhalo masses measured at a
fixed redshift (e.g., \citealt{gao_etal:04, de_lucia_etal:04,
  kravtsov_etal:04, wetzel_etal:09b}), but note that $\msub$ here is
the mass at the time of accretion, the quantity that correlates with
the observational properties of the galaxy contained within
them. Figure \ref{smf} also demonstrates that the subhalo mass
function changes little with redshift. For each bin in $\mhost$, the
N-body data turn over at low mass ratios. This is due to the finite
particle number in the subhalo catalogs utilized; subhalos of the same
initial mass will experience differential tidal forces. This causes
some subhalos to drop below the particle limit of the subhalo catalog,
thus the downturn is artificial. The region where the curves overlap
is the true subhalo mass function. We use a fitting function of the
form

\begin{equation}
\label{e.smf}
\frac{dN}{d\ln\msub} =0.13 \left(\frac{\msub}{\mhost}\right)^{-0.7}\exp\left[-9.9\left(\frac{\msub}{\mhost}\right)^{2.5}\right]
\end{equation}

\noindent when calculating equation (\ref{e.nprev}). 

Figure \ref{tsat} shows the age distribution of subhalos at three
different redshifts. The $y$-axis is the cumulative distribution of
satellites sorted by the time since they were accreted. The value on
the $y$-axis indicates the fraction of subhalos that were accreted
more than time $t$ ago at each redshift. We define $\tsat$ as
$t(z)-t({\rm accretion})$, and $t({\rm accretion})$ is defined as the
time at which a subhalo is first linked to its parent halo by the FOF
halo finder. The shape of $P(\tsat>t)$ clearly depends on redshift; at
earlier epochs, halo densities are higher by $(1+z)^3$, thus the
dynamical timescale decreases by $(1+z)^{-1.5}$. Once the quenched
fraction of satellites is determined, the quenching timescale,$\tq$
found by smoothly interpolating between the numerical data in the
$P(\tsat>t)$ distribution. We determine $\tq$ at the median redshift
of each sample. In tests with the UDS sample, we find negligible
differences between calculating $\tq$ at the median redshift and using
the full $N(z)$ of the sample.

\begin{figure}
\epsscale{1.0} 
\plotone{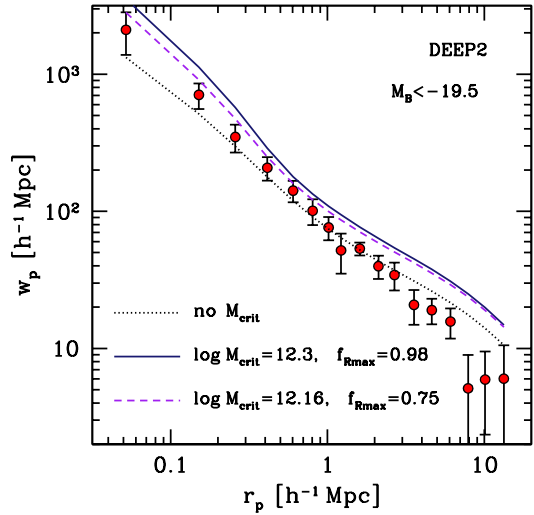}
\caption{ \label{wp_mcrit} The effect of imparting a critical mass
  scale of $\frcen$ for the faint red DEEP2 sample. The dotted curve is
  the model fit from Figure \ref{wp_deep_19.5}. The solid curve is the
  resulting correlation function from a model in which red galaxies
  rapidly dominate the central galaxy population above the critical
  mass obtained in the shock-heating model of
  \cite{dekel_birnboim:06}. The value of $\mredmblue$ is 10 in this
  model. The dashed curve is a less extreme model in which the
  majority of galaxies are red above the critical mass. For this
  model, $\mredmblue=5$. }
\end{figure}

\subsection{Results}

Figure \ref{tquench} shows the results of the model described in \S
5.2 for all four samples (middle four points), with $2\sigma$ error
bars shown.  All four samples are consistent with a value of $\tq\sim
1.8$ Gyr.  There is no obvious dependence on redshift within these
four samples. Figure \ref{tquench} also shows quenching times from
higher and lower redshift obtained from the literature.  At $z\sim
2.3$, \cite{tinker_etal:09_drg} have analyzed the clustering of
distant red galaxies (DRGs). These are massive galaxies that reside in
halos with $\mmin\approx 10^{12}$ \hmsol. Some fraction of these
galaxies have highly attenuated star formation rates, enough to be
classified as red-and-dead (\citealt{labbe_etal:05, papovich_etal:06,
  kriek_etal:06}). \cite{tinker_etal:09_drg} found that the fraction
of satellite galaxies that must be DRGs is between 0.5 and 1
($2\sigma$), while current estimates put the fraction of DRGs that are
quenched between 10-30\% (\citealt{labbe_etal:05,
  papovich_etal:06}). Using these errors and the model described
above, the quenching time for these galaxies is between 300 and 950
Myr (2$\sigma$).

At low redshift, current estimates for the quenching timescale of
satellite galaxies are much longer. \cite{wang_etal:07} used the
clustering of star-forming and quiescent galaxies, as defined by the
$D_n4000$ break in the galaxy spectrum (e.g.,
\citealt{kauffmann_etal:03}), to determine the quenching timescale of
satellite galaxies in the SDSS. They found an $e$-folding time for the
star formation rate of satellite galaxies of 2.5 Gyr. Setting the
quenching time to be the time at which a galaxy's star formation rate
is attenuated by a factor of 10, we find $\tq=5.75$ Gyr. This factor of
10 is roughly the difference between the star-forming and quiescent
samples in the UDS (W09), as well as the blue and red subsamples in
DEEP2 (\citealt{noeske_etal:07a}).  Although \cite{wang_etal:06}
provide no formal error estimate on this number, a rough error bar is
25\%.

The blue circle at $z=0.1$ is the quenching time we determine through
the clustering analysis of \cite{zehavi_etal:05}, who use HOD modeling
of color-selected samples in the SDSS to determine the red satellite
fraction and the red central fraction of galaxies. For galaxies in
magnitude bins centered on $M_r=-19.5$ and $M_r=-20.5$, the model in
section \S 5.1 yields quenching times of $\sim 4$ Gyr. This estimate
is meant to be a consistency check on the \cite{wang_etal:06} results;
a more thorough analysis of the final data release of SDSS galaxies
using these techniques will be presented by Wetzel et.~al. (in
preparation). The green circle at $z=1.4$ is the quenching time
obtained if we impose a prior of $\mredmblue\le 1.1$ on the MCMC
analysis of the UDS clustering. Such a prior increases the fraction of
satellites that are red in order to match the clustering amplitude of
red galaxies, thus decreasing the quenching time.

The shaded band in Figure \ref{tquench} is a power-law that goes as
$(1+z)^{-1.5}$, following the redshift dependence of the dynamical
time of dark matter halos. From $z=0$ to $z=2.3$, the results are
consistent with a quenching timescale that is proportional both to the
dark matter halo dynamical time and the satellite infall time.

\begin{figure*}
\epsscale{1.2} 
\plotone{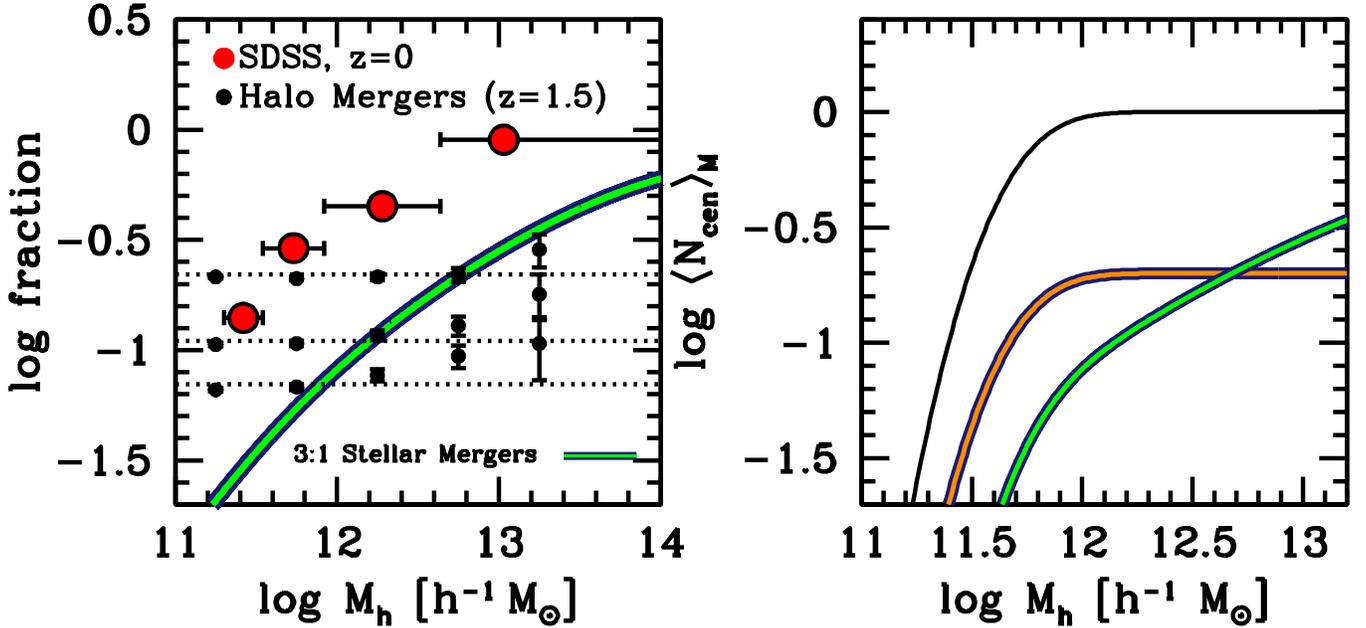}
\vspace{-10.0cm}
\caption{ \label{merger_frac} {\it Left Panel}: Small circles with
  errors represent the fraction of halos that have experienced a
  halo-subhalo (central-satellite) merger within the last Gyr with
  mass ratios smaller than 5:1 ({\it upper}), 3:1 ({\it middle}), and
  2:1 ({\it lower}). The dotted lines show the mean values as for each
  mass ratio limit. The thick solid curve shows the fraction of halos
  whose central {\it galaxy} has experienced a merger with a stellar
  mass ratio smaller than 3:1.  The large red circles represent the
  fraction of red central galaxies from \cite{zehavi_etal:05} analysis
  of SDSS galaxy clustering. {\it Right Panel}: Central occupation
  functions. The thin solid curve represents $\ncen$ for the full UDS
  sample. The thick red line represents a model with
  $\mredmblue=1$. The thick green line represents $\ncen$ produced by
  the merger scenario.}
\end{figure*}

\section{Discussion}

\subsection{How do Central Galaxies Arrive on the Red Sequence?}

The majority of galaxies in the universe are central galaxies. As
discussed in the introduction, current theories on the buildup of the
red sequence focus on how to transform these central galaxies from
star-forming objects to quiescent, red-and-dead objects. One set of
ideas posits that there is a critical halo mass at which star
formation is quenched, either from AGN heating or through sustainable
shock heating (eg, \citealt{croton_etal:06a, dekel_birnboim:06}). This
theory of a critical halo mass is clearly at odds with the clustering
results analyzed here. If there exists a sharp critical mass scale
above which galaxies become red, the relative clustering of red and
blue galaxies would be significantly higher than that observed.

Figure \ref{wp_mcrit} demonstrates this point quantitatively. The
clustering data for the red galaxies in the faint DEEP2 sample are
shown with the best-fit model from Figure \ref{wp_deep_19.5}. The
solid curve is a model in which the fraction of red central galaxies
transitions from 0 to 1 sharply at $\mcrit=2\times10^{12}$ \hmsol,
yielding $\mredmblue=10$. The shape of $\frcen$ is similar to the
dashed curve from Figure \ref{ncen_basic}. To keep the number density
fixed at the observed number density of red galaxies, $\frsat$ is
reduced from 0.6 to 0.45. But even with the lower satellite fraction
in this new model, the amplitude of clustering is significantly higher
than the data. The dashed curve shows the results for a less extreme
model with a lower $\mcrit$ and $\frmax=0.75$, yielding
$\mredmblue=5$. This model is in better agreement with the data in the
one-halo term, but is just as discrepant in the two-halo term.  This
type of model also fails for the bright DEEP2 sample as well as the
COMBO-17 data. For the UDS, a hard critical threshold is the best-fit
model (cf, Figure \ref{wtheta_uds}), but the errors are too large to
distinguish between this model and one in which $\mredmblue=1$.

The `secular' scenario (\citealt{bower_etal:06}), where instead of a
critical halo mass threshold, there is a critical {\it galaxy} mass
threshold, also predicts a strong dependence of $\frcen$ on halo
mass. At $M=10^{11.3}$ \hmsol\ $\frcen\sim 0$, and by $M=10^{12.0}$
\hmsol\ the fraction of red central galaxies is near unity (see Figure
6 in \citealt{hopkins_etal:08b}). The merger scenario has a weaker
dependence of $\frcen$ on $M$, but still produces a high $\mredmblue$
ratio. The {\it halo} merger rate depends little on halo mass
(\citealt{cohn_etal:01, fakhouri_ma:07}). This is also true of the
merger rate of subhalos with host halos, which should more accurately
track the galaxy merger rate (\citealt{wetzel_etal:09b}). However, the
mapping between halo mass and galaxy mass is complex; the $\mgal/M$
ratio peaks at $M \sim 10^{12}$ \hmsol\ and falls off as a power law
at higher and lower halo halo masses (see, eg, \citealt{wang_etal:07,
  conroy_wechsler:09, moster_etal:09}). Because $\mgal \sim M^{0.4}$
at $M\gtrsim 10^{12}$ \hmsol, a 10:1 halo-halo minor merger can
contain galaxies with mass ratios of 3:1, yielding a major galaxy
merger (\citealt{hopkins_etal:08b, maller:08}). In all calculations
below we use the redshift-dependent stellar mass functions of
\cite{marchesini_etal:07}  assign stellar masses to halos and
  subhalos using abundance matching, interpolating between redshifts
to obtain the stellar mass function at the redshift of a given sample.

Figure \ref{merger_frac} shows the $z=1.5$ red central fraction
induced by major mergers, which we define as stellar mass ratios of
3:1 or closer, that have occured over the previous Gyr. The small
circles indicate the fraction of {\it halos} that have experienced a
halo-subhalo (central-satellite) merger with mass ratios smaller than
2:1, 3:1, and 5:1, respectively, from the simulations of
\cite{wetzel_etal:09b}. The lack of strong dependence on halo mass is
clear.  Using abundance matching to map halo mass onto galaxy mass, a
3:1 mass ratio for {\it galaxies} in those same halos yields a red
central fraction of $\sim 10\%$ at $M=10^{12}$ \hmsol, rising to $\sim
30\%$ at $10^{13}$ \hmsol\ and $\sim 60\%$ at $10^{14}$ \hmsol. The
right hand panel of Figure \ref{merger_frac} compares the central
occupation function for red galaxies for model with $\mredmblue=1$ to
$\ncen$ created by the merger scenario. At $z=1.5$, the merger
scenario yields a central galaxy number density that is too low by
roughly a factor of 3. As stated above, the halo merger rates are
calculated using a lookback time of 1 Gyr. Increasing this timescale
increases the fraction of halos that have experienced mergers, but it
does not change the {\it shape} of the merged fraction with halo mass,
thus we consider the amplitude of this function to be a free parameter
set by the overall number of red galaxies. Using the 1 Gyr lookback time
to calculate $\fm$, the merger scenario yields
$\mredmblue=2.2$. Increasing the normalization of $\fm$ to produce the
correct number density of yields $\mredmblue=2.9$. These values are in
agreement with the weak constraints on $\frcen$ shown in Figure
\ref{mcen_fred}. At $z=0.8$, the redshift of the faint DEEP2 sample,
the merger scenario yields $\mredmblue=3.8$ using the 1 Gyr lookback
time, and it yields $\mredmblue=5.2$ if the amplitude of $\fm$ is
increased to produce the proper red central number density. The higher
values reflect the change in the halo mass function; under the merger
scenario most high-mass halos are quenched, and the abundance of
high-mass halos has increased substantively from $z=1.4$ to $z=0.8$.
Results for COMBO-17 are similar. These values are strongly excluded
for both the DEEP2 and COMBO-17 results.  Thus, while the UDS data are
broadly consistent with the merger scenario, DEEP2 and COMBO-17 are
not.

To connect with the low redshift universe, Figure \ref{merger_frac}
also shows the dependence of $\frcen$ on halo mass obtained from SDSS
clustering by \cite{zehavi_etal:05}. The shape of $\frcen$ at $z=0.1$
is consistent with the merger scenario (see also
\citealt{hopkins_etal:08b}). If the COMBO-17 results are
representative of the clustering of blue and red galaxies at $z=0.6$,
what occurs from $z=0.6$ to $z=0$ to induce the change in $\frcen$?
Selection effects seem unlikely to impart a mass dependence in
$\frcen$ in SDSS; indeed, the results of \cite{maller_etal:09}
demonstrate that many low-luminosity galaxies in SDSS are
misclassified as red due to dust effects. Similarly, selection biases
seem unlikely to {\it remove} a mass dependence in $\frcen$ in
COMBO-17 and DEEP2. Red galaxies are selected by different methods in
these two samples; for COMBO-17, spectral energy distribution fitting
is used; in DEEP2, a color cut similar to SDSS is employed.

Is there evolution in the mechanism by which central galaxies quench
their star formation? For halos of $\lesssim 10^{11.5}$ \hmsol, the
red central fraction for the high-redshift samples is similar to $z=0$
estimates, modulo the differences in the classification of red
galaxies between the various surveys. But $\frcen$ at
$M\sim 10^{12-12.5}$ \hmsol\ must increase by nearly a factor of two,
and this fraction must increase by nearly a factor of three at
$M\gtrsim 10^{13}$ \hmsol. More time elapses from $z=0.6$ to $z=0$
(5.8 Gyr) than elapses from $z=1.4$ to $z=0.6$ (3.3 Gyr). Although
halo dynamical times evolve strongly with redshift, there are roughly
the same number of dynamical times between SDSS and COMBO-17 as there
are between COMBO-17 and the UDS (approximately 6). Thus evolution
should not be ruled out as a possibility.

\subsection{The Role of Satellite Galaxies in Building up the Red Sequence}

\begin{figure}
\epsscale{1.} 
\plotone{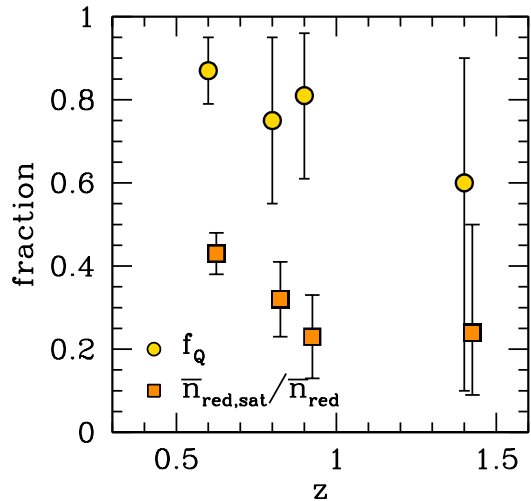}
\caption{ \label{fractions} Circles indicate the fraction of red
  satellite galaxies that were quenched after accretion onto their
  parent halo, $\fq$, as defined by equation (\ref{e.fq}). Squares
  indicate the fraction of the red sequence made up of satellite
  galaxies at the magnitude threshold for each sample. The redshifts
  for the squares are slightly offset for clarity. All errorbars are
  2$\sigma$. }
\end{figure}

At all redshifts, from $z=2.3$ to $z=0$, the fraction of satellites
that are red is roughly consistent with $\sim 60\%$ of all satellites
in the sample. This is more than double the fraction of centrals which
are red, demonstrating that satellites are not simply made red before
infall. Figure \ref{fractions} show what fraction of the red sequence
is made up satellite galaxies for each sample, as well as the fraction
of those satellites that were quenched after becoming satellites. At
redshifts less than $z=2$, 60\% of satellites being red translates to
$\sim 30\%$ of the red sequence being made of up satellites. Roughly
70-80\% of red satellites became red {\it after} accretion, in
agreement with results for low-mass galaxies at $z=0$
(\citealt{vdb_etal:08}), although the UDS results are consistent with
a value as low of 10\%. Thus satellite galaxies play a major role in
building up the red sequence. The surprising fact is that these
numbers do not vary substantially with redshift. The survival time of
a subhalo scales with the dynamical time of the host halo,
$(1+z)^{-1.5}$. Thus, to keep $\frsat$ a constant requires the same
redshift dependence in the quenching time, as shown in Figure
\ref{tquench}. Given the finite lifetimes of subhalos, satellite
galaxies observed at various redshifts are distinct sets of
galaxies. The fraction of subhalos at $z=0.6$ that survive to $z=0$ is
$\sim 20\%$. The fraction from $z=0.9$ that are extant today is
less than $5\%$. Thus the satellite galaxies analyzed in the SDSS by
\cite{vdb_etal:08} are not the same as those quantified here, yet
their statistics are quite similar.

\cite{wang_etal:07} used clustering to constrain the quenching time at
$z=0$ (cf, Figure \ref{tquench}). They posit that this quenching time
does not evolve with redshift, and that this fixed $\tq$ value yields
a color-density relation that largely goes away at $z\ge 1$, in
qualitative agreement with the observations of
\cite{cooper_etal:07}. However, a quenching time that is fixed at 5
Gyr would be incompatible with all of the correlation functions
analyzed in this paper. From Figure \ref{tsat}, it can be seen that
{\it no} satellites are older than 6 Gyr at $z\gtrsim 0.6$, thus there
would be no satellites quenched after infall at $z\gtrsim 0.6$.  The
satellite red fraction would simply trace the central red fraction,
and the clustering of red galaxies would be the same as blue galaxies.
The \cite{cooper_etal:07} measurements were made using the DEEP2
survey, thus they must be consistent with the correlation functions of
\cite{coil_etal:08} from the same data. The lack of a strong
color-density relation, as well as its gradual attenuation with
increasing redshift, are compatible with the HOD results obtained
here; if red central galaxies are a random sample of all central
galaxies, then they produce no color-density relation. The high
fraction of satellite galaxies that are red would induce a correlation
between color and density, but as redshift increases the overall
fraction of galaxies that are satellites decreases. Thus $\frsat$ can
remain fixed in redshift and still result in a decreasing satellite
fraction within the red sequence. It should be noted, however, that
the high relative bias between red and blue galaxies at $1<z<2$ in W09
implies that the color-density relation does exist in the
UDS. \cite{quadri_etal:07} find the same result at higher redshift in
the MUSYC survey. Further data are required to resolve this
discrepancy.

The simple model for obtaining the satellite quenching time in \S 5.1
neglects certain effects. The quenching time may depend on both the
age of a satellite and the details of its orbit (as suggested by
\citealt{balogh_etal:09}), as well as the mass of the subhalo or the
stellar mass of the satellite galaxy. Another explicit assumption in
this model is that the quenching can only begin after a galaxy passes
inside the virial radius of the parent halo. These questions will be
addressed in a subsequent paper (Wetzel et.~al., in preparation).

\subsection{Inferences from the Luminosity Function}

At redshift zero, our now-standard picture is that the fraction of
central galaxies that are red increases with halo mass. Above
$M\approx 10^{13.5}$ \hmsol, nearly all central galaxies are red
(\citealt{zehavi_etal:05, weinmann_etal:06, yang_etal:08}). This
scenario is influenced by measurements of the galaxy luminosity
function; red fraction and galaxy luminosity are strongly correlated
(e.g., \citealt{blanton_etal:03} others). The results at higher
redshift, however, are quite different. Above $M_B=-19.5$, the
abundance of red galaxies in DEEP2 has little correlation with
luminosity (\citealt{willmer_etal:06}). In Figure
\ref{number_densities}, the red fraction increases by only 9\% between
the faint and bright DEEP2 samples. At lower luminosities, the
fraction of red galaxies does decrease. Thus, the value of
$\mredmblue$ should increase at lower luminosity thresholds. If there
is a preferential halo mass scale for star formation quenching in
central galaxies, it is below the mass scale probed by the samples we
have analyzed, $M\lesssim 10^{11}$ \hmsol.

\subsection{Possible Systematic Errors}

A concern in this analysis is the use of diagonal error bars only, as
opposed to using the full covariance matrix for each sample. The
clustering statistics used here are known to be correlated in the
two-halo term and largely uncorrelated in the one-halo term (see, eg,
figures in \citealt{zehavi_etal:05, phleps_etal:06, blake_etal:08,
  tinker_etal:09_drg}). Thus, the use of diagonal errors over-weights
the contribution of the data in the two-halo term to the $\chi^2$ of
each fit. Using our 720 \hmpc\ simulation, we have made mock galaxy
distributions that roughly match the faint DEEP2 sample. The mass
resolution in this simulation is not enough to model $M\lesssim
10^{11.3}$ \hmsol\ halos, thus we are forced to modify $\ncen$ of the
best-fit model to obtain the proper number density of galaxies. We
divide the simulation volume into $6^3=216$ cubic subvolumes of 120
\hmpc\ per side. We calculate $\wp$ within each subvolume using the
Landy-Szalay estimator and accounting for the variation in the number
density of galaxies in each subvolume. We are not able to obtain a
robust covariance matrix for both the red and blue subsamples, thus we
adopt the same (normalized) covariance matrix for both subsamples. We
multiply each element by the diagonal elements measured by
\cite{coil_etal:08}, ie $C_{ij} = \hat{C}_{ij}e_ie_j$, where
$\hat{C}_{ij}$ is the normalized matrix from the mocks,
$\hat{C}_{ij}=C_{ij}/\sqrt{C_{ii}C_{jj}}$. Although the covariance
matrices of the red and blue subsamples should vary in detail, our
test replicates the general consequences of a covariance matrix: some
correlation of data in the one-halo term, with an increasing amplitude
of off-diagonal terms in the two-halo term. We find virtually no
difference in the $\mredmblue$-$\frsat$ constraints when using these
test matrices. We conclude that our exact answers may vary slightly
with use of proper estimates of $C_{ij}$, but the results will not
change enough to alter our conclusions.

Selection effects are also a concern. If $\frcen$ did depend on halo
mass, either through a critical mass scale or a weaker dependence from
a merger-like scenario, dust contamination would tend to mask this
mass dependence. If galaxies are misclassified as red or quenched due
to dust contamination, the large-scale amplitude of red galaxy
clustering is attenuated. In terms of the HOD models presented here,
and under the assumption that $\frcen$ was actually monotonically
increasing with halo mass, dust would take low-mass blue central
galaxies and place then in the red sample, reducing
$\mredmblue$. However, this effect seems unlikely to produce $\frcen$
in the best-fit DEEP2 model. First, if $\frcen$ had a strong mass
dependence to it, such that the fraction of red central galaxies at
$M\gtrsim 10^{13}$ \hmsol\ were $\gtrsim 75\%$ (a value taken from the
analysis of $z=0$ clustering in the SDSS by \citealt{zehavi_etal:05}),
then a mass-independent dust contamination would smooth our $\frcen$
but not be able to remove this trend completely. More importantly, the
low values of $\frcen$ at high halo masses in the DEEP2 models ($\sim
0.15-0.25$) are at odds with this scenario.

One degree of freedom we have not introduced into the HOD model used
here is a mass-dependence on the fraction of satellites that are
red. We have performed tests with the faint DEEP2 data, introducing a
mass scale below which $\frsat$ is attenuated with a Gaussian
cutoff. The constraints on this cutoff mass are weak and do not change
either $\tq$ or $\mredmblue$. The color-dependent HOD in
\cite{zehavi_etal:05} contained a mass-dependent parameterization in
$\frsat$, but in most clustering samples
$\frsat$ was found to be constant. If red satellite galaxies prefer
higher mass halos, the red correlation function would contain a more
pronounced break between the 1-halo and 2-halo terms, and the shape of
the red correlation function at small scales would not be a
power-law. In the DEEP2 data, $\wp$ for the red galaxies is a power
law down to $r_p\sim 50$ \hkpc, and it does not contain a strong break
at $r_p\sim 1$ \hmpc.

Although $\asat=1$ is a well-motivated prior for luminosity
thresholds, assuming this power-law index for color subsamples is less
cut and dry. A better fit to the faint DEEP2 data is obtained if I
parameterize $\nsat$ for blue and red subsamples as independent
power-law functions, with different values of $\msat$ and $\asat$ for
each. The $\chi^2$ for this model is 38.6, a significant improvement
on the fit from the fiducial model. However, to produce this $\chi^2$,
both $\asat$ for the red and blue galaxies is well below unity. If
$\asat=1$ for the full sample, then a value of $\asat<1$ for blue
galaxies requires $\asat>1$ for the red subsample (or vice
versa). Models that produce a better fit than the fiducial model have
both values below unity, with the best-fit model yielding $\asat=0.39$
for the red subsample and $\asat=0.48$ for the blue subsample. These
values are firmly excluded from theoretical expectations and
observational data (cf. \S 3.1). Within the range $0.9\le \asat \le
1.1$, our constraints in Figure \ref{mcen_fred} are unchanged.

\subsection{Comparison to Results from the NDWFS}

\cite{brown_etal:08} perform HOD analysis on the clustering of red
galaxies from $z=0.2-1.0$ in the NDWFS imaging survey
(\citealt{ndwfs}). At $\sim 7$ square degrees, this sample is much
larger than all the samples analyzed in this paper. Additionally,
\cite{brown_etal:08} use the full covariance for all halo occupation
analysis. They find that the $\frcen=0.5$ at $10^{11.9}$ \hmsol,
increasing with halo mass. However, the HOD model for central galaxies
used in their analysis is identical to Equation (\ref{e.ncen})
here. Essentially, they model red galaxy samples with an HOD
constructed for a full galaxy sample. This means that $\frcen$ must
increase with mass and it will approach unity at high halos masses by
construction. Without a corresponding sample of blue galaxies, it is
difficult to constrain the asymptotic fraction of red central
galaxies. Because \cite{brown_etal:08} only have photometric redshifts
of the galaxies in their sample, they can only measure $\wth$, which
has larger fractional errors at large scales relative to measurements
of $\wp$ over similar volumes. All HOD fits to their data are
somewhat high in the two-halo term, which could be caused by the
assumption that $\frcen$ increases to unity (cf, Figure
\ref{wp_mcrit}). However, the errors on $\wth$ are large enough such
that they are all statistically good fits. Corresponding measurements
of the clustering of blue galaxies in the same field would help
resolve the discrepancy between the results of \cite{brown_etal:08}
and those in this paper.

\section{Summary}

In this paper we have analyzed clustering measurements of red and blue
galaxies at four different redshifts, $z=1.4$, $z=0.9$, $z=0.8$, and
$z=0.6$. We have used the halo occupation distribution to determine the
fraction galaxies are that red, broken into contributions of red
central galaxies and red satellite galaxies. In all the samples
analyzed, the higher clustering amplitude of red galaxies can be fully
accounted for the the high fraction of satellite galaxies that are
red; on average, the overall red fraction of each sample is $\sim
25\%$, while the fraction of satellites that are red is $\sim
60\%$. Thus the fraction of the red sequence that is satellite
galaxies is $\sim 30\%$. The high value of $\frsat$ produces enhanced
clustering at both large and small scales relative to blue galaxies
above the same magnitude threshold. Imparting a halo mass dependence
on the fraction of central galaxies that are red enhances the relative
bias between red and blue subsamples are large scales. The data are
not compatible with such a scenario, favoring $\mredmblue=1$ for
central galaxies.

From galaxy-halo mass-to-light ratios (and stellar mass to halo mass
ratios), it is inferred that galaxy formation efficiency is maximal at
$M\sim 10^{12}$ \hmsol\ (\citealt{yang_etal:03, vdb_etal:07,
  tinker_etal:05, conroy_wechsler:09, moster_etal:09}). This mass
scale increases slowly with redshift out to $z\sim 1$
(\citealt{conroy_wechsler:09, moster_etal:09}). This implies that a
central galaxy's integrated star formation history is dependent on
halo mass. This leads to a correlation between instantaneous star
formation rate and halo mass (see Figure 8 in
\citealt{conroy_wechsler:09}). In contrast, our results imply that the
fraction of galaxies that are quenched is independent of halo mass at
$z\gtrsim 0.6$.

Models in which central galaxies become red at a critical halo mass
scale of $\sim 10^{12}$ \hmsol\ are strongly excluded by
the data available at $z\lesssim 1$. A scenario in which mergers
produce red galaxies has a weaker dependence of red central fraction
on halo mass, but these models also produce values of $\mredmblue$
larger than unity, values that outside the 2-$\sigma$ constraints from
DEEP2 and COMBO-17.

The constant value of $\frsat$ implies that the quenching timescale
for satellite galaxies depends on redshift in the same manner as the
dynamical timescale of dark matter halos. Given the errors, $\tq$ for
all four samples is consistent with $\tq=1.8$ Gyr. Current estimates
of $\tq$ at $z=0$ are much longer, $\sim 5$ Gyr, while preliminary
estimates of $\tq$ at $z>2$ are less than a Gyr.

Given the small volume of each sample, more data are required to
ameliorate the possibility that sample variance is significantly
influencing our results. Additionally, clustering measurements at
redshift between 0 and 0.6 can shed light on discrepancies seen
between $z=0$ and the samples analyzed here. Upcoming results from the
Galaxy and Mass Assembly survey (GAMA; \citealt{driver_etal:09}), as
well as the clustering of blue galaxies from the NDWFS, should show
intermediate stages of evolution in both the quenching mechanism for
central galaxies and the quenching timescale for satellites.


\bibliography{../risa}

\begin{thebibliography}{116}
\expandafter\ifx\csname natexlab\endcsname\relax\def\natexlab#1{#1}\fi

\bibitem[{{Abadi} {et~al.}(1999){Abadi}, {Moore}, \& {Bower}}]{abadi_etal:99}
{Abadi}, M.~G., {Moore}, B., \& {Bower}, R.~G. 1999, \mnras, 308, 947

\bibitem[{{Balogh} {et~al.}(2009){Balogh}, {McGee}, {Wilman}, {Bower}, {Hau},
  {Morris}, {Mulchaey}, {Oemler}, {Parker}, \& {Gwyn}}]{balogh_etal:09}
{Balogh}, M.~L., {McGee}, S.~L., {Wilman}, D., {Bower}, R.~G., {Hau}, G.,
  {Morris}, S.~L., {Mulchaey}, J.~S., {Oemler}, Jr., A., {Parker}, L., \&
  {Gwyn}, S. 2009, ArXiv e-prints

\bibitem[{{Bell} {et~al.}(2004){Bell}, {Wolf}, {Meisenheimer}, {Rix}, {Borch},
  {Dye}, {Kleinheinrich}, {Wisotzki}, \& {McIntosh}}]{bell_etal:04}
{Bell}, E.~F., {Wolf}, C., {Meisenheimer}, K., {Rix}, H.-W., {Borch}, A.,
  {Dye}, S., {Kleinheinrich}, M., {Wisotzki}, L., \& {McIntosh}, D.~H. 2004,
  \apj, 608, 752

\bibitem[{{Berlind} \& {Weinberg}(2002)}]{berlind_weinberg:02}
{Berlind}, A.~A. \& {Weinberg}, D.~H. 2002, \apj, 575, 587

\bibitem[{{Berlind} {et~al.}(2003){Berlind}, {Weinberg}, {Benson}, {Baugh},
  {Cole}, {Dav{\' e}}, {Frenk}, {Jenkins}, {Katz}, \&
  {Lacey}}]{berlind_etal:03}
{Berlind}, A.~A., {Weinberg}, D.~H., {Benson}, A.~J., {Baugh}, C.~M., {Cole},
  S., {Dav{\' e}}, R., {Frenk}, C.~S., {Jenkins}, A., {Katz}, N., \& {Lacey},
  C.~G. 2003, \apj, 593, 1

\bibitem[{{Birnboim} \& {Dekel}(2003)}]{birnboim_dekel:03}
{Birnboim}, Y. \& {Dekel}, A. 2003, \mnras, 345, 349

\bibitem[{{Blake} {et~al.}(2008){Blake}, {Collister}, \&
  {Lahav}}]{blake_etal:08}
{Blake}, C., {Collister}, A., \& {Lahav}, O. 2008, \mnras, 385, 1257

\bibitem[{{Blanton} {et~al.}(2003{\natexlab{a}}){Blanton}, {Hogg}, {Bahcall},
  {Baldry}, {Brinkmann}, {Csabai}, {Eisenstein}, {Fukugita}, {Gunn},
  {Ivezi{\'c}}, {Lamb}, {Lupton}, {Loveday}, {Munn}, {Nichol}, {Okamura},
  {Schlegel}, {Shimasaku}, {Strauss}, {Vogeley}, \&
  {Weinberg}}]{blanton_etal:03cmd}
{Blanton}, M.~R., {Hogg}, D.~W., {Bahcall}, N.~A., {Baldry}, I.~K.,
  {Brinkmann}, J., {Csabai}, I., {Eisenstein}, D., {Fukugita}, M., {Gunn},
  J.~E., {Ivezi{\'c}}, {\v Z}., {Lamb}, D.~Q., {Lupton}, R.~H., {Loveday}, J.,
  {Munn}, J.~A., {Nichol}, R.~C., {Okamura}, S., {Schlegel}, D.~J.,
  {Shimasaku}, K., {Strauss}, M.~A., {Vogeley}, M.~S., \& {Weinberg}, D.~H.
  2003{\natexlab{a}}, \apj, 594, 186

\bibitem[{{Blanton} {et~al.}(2003{\natexlab{b}}){Blanton}, {Hogg}, {Bahcall},
  {Brinkmann}, {Britton}, {Connolly}, {Csabai}, {Fukugita}, {Loveday},
  {Meiksin}, {Munn}, {Nichol}, {Okamura}, {Quinn}, {Schneider}, {Shimasaku},
  {Strauss}, {Tegmark}, {Vogeley}, \& {Weinberg}}]{blanton_etal:03}
{Blanton}, M.~R., {Hogg}, D.~W., {Bahcall}, N.~A., {Brinkmann}, J., {Britton},
  M., {Connolly}, A.~J., {Csabai}, I., {Fukugita}, M., {Loveday}, J.,
  {Meiksin}, A., {Munn}, J.~A., {Nichol}, R.~C., {Okamura}, S., {Quinn}, T.,
  {Schneider}, D.~P., {Shimasaku}, K., {Strauss}, M.~A., {Tegmark}, M.,
  {Vogeley}, M.~S., \& {Weinberg}, D.~H. 2003{\natexlab{b}}, \apj, 592, 819

\bibitem[{{Bower} {et~al.}(2006){Bower}, {Benson}, {Malbon}, {Helly}, {Frenk},
  {Baugh}, {Cole}, \& {Lacey}}]{bower_etal:06}
{Bower}, R.~G., {Benson}, A.~J., {Malbon}, R., {Helly}, J.~C., {Frenk}, C.~S.,
  {Baugh}, C.~M., {Cole}, S., \& {Lacey}, C.~G. 2006, \mnras, 370, 645

\bibitem[{{Brown} {et~al.}(2008){Brown}, {Zheng}, {White}, {Dey}, {Jannuzi},
  {Benson}, {Brand}, {Brodwin}, \& {Croton}}]{brown_etal:08}
{Brown}, M.~J.~I., {Zheng}, Z., {White}, M., {Dey}, A., {Jannuzi}, B.~T.,
  {Benson}, A.~J., {Brand}, K., {Brodwin}, M., \& {Croton}, D.~J. 2008, \apj,
  682, 937

\bibitem[{{Butcher} \& {Oemler}(1978)}]{butcher_oemler:78}
{Butcher}, H. \& {Oemler}, Jr., A. 1978, \apj, 226, 559

\bibitem[{{Butcher} \& {Oemler}(1984)}]{butcher_oemler:84}
---. 1984, \apj, 285, 426

\bibitem[{{Cattaneo} {et~al.}(2006){Cattaneo}, {Dekel}, {Devriendt},
  {Guiderdoni}, \& {Blaizot}}]{cattaneo_etal:06}
{Cattaneo}, A., {Dekel}, A., {Devriendt}, J., {Guiderdoni}, B., \& {Blaizot},
  J. 2006, \mnras, 370, 1651

\bibitem[{{Cohn} {et~al.}(2001){Cohn}, {Bagla}, \& {White}}]{cohn_etal:01}
{Cohn}, J.~D., {Bagla}, J.~S., \& {White}, M. 2001, \mnras, 325, 1053

\bibitem[{{Coil} {et~al.}(2008){Coil}, {Newman}, {Croton}, {Cooper}, {Davis},
  {Faber}, {Gerke}, {Koo}, {Padmanabhan}, {Wechsler}, \&
  {Weiner}}]{coil_etal:08}
{Coil}, A.~L., {Newman}, J.~A., {Croton}, D., {Cooper}, M.~C., {Davis}, M.,
  {Faber}, S.~M., {Gerke}, B.~F., {Koo}, D.~C., {Padmanabhan}, N., {Wechsler},
  R.~H., \& {Weiner}, B.~J. 2008, \apj, 672, 153

\bibitem[{{Collister} \& {Lahav}(2005)}]{collister_lahav:05}
{Collister}, A.~A. \& {Lahav}, O. 2005, \mnras, 361, 415

\bibitem[{{Conroy} \& {Wechsler}(2009)}]{conroy_wechsler:09}
{Conroy}, C. \& {Wechsler}, R.~H. 2009, \apj, 696, 620

\bibitem[{{Conroy} {et~al.}(2006){Conroy}, {Wechsler}, \&
  {Kravtsov}}]{conroy_etal:06}
{Conroy}, C., {Wechsler}, R.~H., \& {Kravtsov}, A.~V. 2006, \apj, 647, 201

\bibitem[{{Cooper} {et~al.}(2007){Cooper}, {Newman}, {Coil}, {Croton}, {Gerke},
  {Yan}, {Davis}, {Faber}, {Guhathakurta}, {Koo}, {Weiner}, \&
  {Willmer}}]{cooper_etal:07}
{Cooper}, M.~C., {Newman}, J.~A., {Coil}, A.~L., {Croton}, D.~J., {Gerke},
  B.~F., {Yan}, R., {Davis}, M., {Faber}, S.~M., {Guhathakurta}, P., {Koo},
  D.~C., {Weiner}, B.~J., \& {Willmer}, C.~N.~A. 2007, \mnras, 376, 1445

\bibitem[{{Cooper} {et~al.}(2006){Cooper}, {Newman}, {Croton}, {Weiner},
  {Willmer}, {Gerke}, {Madgwick}, {Faber}, {Davis}, {Coil}, {Finkbeiner},
  {Guhathakurta}, \& {Koo}}]{cooper_etal:06}
{Cooper}, M.~C., {Newman}, J.~A., {Croton}, D.~J., {Weiner}, B.~J., {Willmer},
  C.~N.~A., {Gerke}, B.~F., {Madgwick}, D.~S., {Faber}, S.~M., {Davis}, M.,
  {Coil}, A.~L., {Finkbeiner}, D.~P., {Guhathakurta}, P., \& {Koo}, D.~C. 2006,
  \mnras, 370, 198

\bibitem[{{Cooray} \& {Sheth}(2002)}]{cooray_sheth:02}
{Cooray}, A. \& {Sheth}, R. 2002, \physrep, 372, 1

\bibitem[{{Croton} {et~al.}(2006){Croton}, {Springel}, {White}, {De Lucia},
  {Frenk}, {Gao}, {Jenkins}, {Kauffmann}, {Navarro}, \&
  {Yoshida}}]{croton_etal:06a}
{Croton}, D.~J., {Springel}, V., {White}, S.~D.~M., {De Lucia}, G., {Frenk},
  C.~S., {Gao}, L., {Jenkins}, A., {Kauffmann}, G., {Navarro}, J.~F., \&
  {Yoshida}, N. 2006, \mnras, 365, 11

\bibitem[{{Davis} \& {Peebles}(1983)}]{davis_peebles:83}
{Davis}, M. \& {Peebles}, P.~J.~E. 1983, \apj, 267, 465

\bibitem[{{De Lucia} \& {Blaizot}(2007)}]{delucia_blaizot:07}
{De Lucia}, G. \& {Blaizot}, J. 2007, \mnras, 375, 2

\bibitem[{{De Lucia} {et~al.}(2004){De Lucia}, {Kauffmann}, {Springel},
  {White}, {Lanzoni}, {Stoehr}, {Tormen}, \& {Yoshida}}]{de_lucia_etal:04}
{De Lucia}, G., {Kauffmann}, G., {Springel}, V., {White}, S.~D.~M., {Lanzoni},
  B., {Stoehr}, F., {Tormen}, G., \& {Yoshida}, N. 2004, \mnras, 348, 333

\bibitem[{{Dekel} \& {Birnboim}(2006)}]{dekel_birnboim:06}
{Dekel}, A. \& {Birnboim}, Y. 2006, \mnras, 368, 2

\bibitem[{{Dekel} \& {Birnboim}(2008)}]{dekel_birnboim:08}
---. 2008, \mnras, 383, 119

\bibitem[{{Dekel} {et~al.}(2008){Dekel}, {Birnboim}, {Engel}, {Freundlich},
  {Goerdt}, {Mumcuoglu}, {Neistein}, {Pichon}, {Teyssier}, \&
  {Zinger}}]{dekel_etal:08}
{Dekel}, A., {Birnboim}, Y., {Engel}, G., {Freundlich}, J., {Goerdt}, T.,
  {Mumcuoglu}, M., {Neistein}, E., {Pichon}, C., {Teyssier}, R., \& {Zinger},
  E. 2008, Nature, accepted (arXiv:0808.0553)

\bibitem[{{Driver} {et~al.}(2009){Driver}, {the GAMA Team}, {Baldry},
  {Bamford}, {Bland-Hawthorn}, {Bridges}, {Cameron}, {Conselice}, {Couch},
  {Croom}, {Cross}, {Driver}, {Dunne}, {Eales}, {Edmondson}, {Ellis}, {Frenk},
  {Graham}, {Jones}, {Hill}, {Hopkins}, {van Kampen}, {Kuijken}, {Lahav},
  {Liske}, {Loveday}, {Nichol}, {Norberg}, {Oliver}, {Parkinson}, {Peacock},
  {Phillipps}, {Popescu}, {Prescott}, {Proctor}, {Sharp}, {Staveley-Smith},
  {Sutherland}, {Tuffs}, \& {Warren}}]{driver_etal:09}
{Driver}, S.~P., {the GAMA Team}, {Baldry}, I.~K., {Bamford}, S.,
  {Bland-Hawthorn}, J., {Bridges}, T., {Cameron}, E., {Conselice}, C., {Couch},
  W.~J., {Croom}, S., {Cross}, N.~J.~G., {Driver}, S.~P., {Dunne}, L., {Eales},
  S., {Edmondson}, E., {Ellis}, S.~C., {Frenk}, C.~S., {Graham}, A.~W.,
  {Jones}, H., {Hill}, D., {Hopkins}, A., {van Kampen}, E., {Kuijken}, K.,
  {Lahav}, O., {Liske}, J., {Loveday}, J., {Nichol}, B., {Norberg}, P.,
  {Oliver}, S., {Parkinson}, H., {Peacock}, J.~A., {Phillipps}, S., {Popescu},
  C.~C., {Prescott}, M., {Proctor}, R., {Sharp}, R., {Staveley-Smith}, L.,
  {Sutherland}, W., {Tuffs}, R.~J., \& {Warren}, S. 2009, in IAU Symposium,
  Vol. 254, IAU Symposium, ed. J.~{Andersen}, J.~{Bland-Hawthorn}, \&
  B.~{Nordstr{\"o}m}, 469--474

\bibitem[{{Dunkley} {et~al.}(2008){Dunkley}, {Komatsu}, {Nolta}, {Spergel},
  {Larson}, {Hinshaw}, {Page}, {Bennett}, {Gold}, {Jarosik}, {Weiland},
  {Halpern}, {Hill}, {Kogut}, {Limon}, {Meyer}, {Tucker}, {Wollack}, \&
  {Wright}}]{dunkley_etal:08}
{Dunkley}, J., {Komatsu}, E., {Nolta}, M.~R., {Spergel}, D.~N., {Larson}, D.,
  {Hinshaw}, G., {Page}, L., {Bennett}, C.~L., {Gold}, B., {Jarosik}, N.,
  {Weiland}, J.~L., {Halpern}, M., {Hill}, R.~S., {Kogut}, A., {Limon}, M.,
  {Meyer}, S.~S., {Tucker}, G.~S., {Wollack}, E., \& {Wright}, E.~L. 2008,
  ApJS, in press

\bibitem[{{Eke} {et~al.}(2004){Eke}, {Baugh}, {Cole}, {Frenk}, {Norberg},
  {Peacock}, {Baldry}, {Bland-Hawthorn}, {Bridges}, {Cannon}, {Colless},
  {Collins}, {Couch}, {Dalton}, {de Propris}, {Driver}, {Efstathiou}, {Ellis},
  {Glazebrook}, {Jackson}, {Lahav}, {Lewis}, {Lumsden}, {Maddox}, {Madgwick},
  {Peterson}, {Sutherland}, \& {Taylor}}]{eke_etal:04}
{Eke}, V.~R., {Baugh}, C.~M., {Cole}, S., {Frenk}, C.~S., {Norberg}, P.,
  {Peacock}, J.~A., {Baldry}, I.~K., {Bland-Hawthorn}, J., {Bridges}, T.,
  {Cannon}, R., {Colless}, M., {Collins}, C., {Couch}, W., {Dalton}, G., {de
  Propris}, R., {Driver}, S.~P., {Efstathiou}, G., {Ellis}, R.~S.,
  {Glazebrook}, K., {Jackson}, C., {Lahav}, O., {Lewis}, I., {Lumsden}, S.,
  {Maddox}, S., {Madgwick}, D., {Peterson}, B.~A., {Sutherland}, W., \&
  {Taylor}, K. 2004, \mnras, 348, 866

\bibitem[{{Fakhouri} \& {Ma}(2007)}]{fakhouri_ma:07}
{Fakhouri}, O. \& {Ma}, C.-P. 2007, \mnras, submited (ArXiv:0710:4567)

\bibitem[{{Fisher}(1995)}]{fisher:95}
{Fisher}, K.~B. 1995, \apj, 448, 494

\bibitem[{{Gao} {et~al.}(2004){Gao}, {White}, {Jenkins}, {Stoehr}, \&
  {Springel}}]{gao_etal:04}
{Gao}, L., {White}, S.~D.~M., {Jenkins}, A., {Stoehr}, F., \& {Springel}, V.
  2004, \mnras, 355, 819

\bibitem[{{Gerke} {et~al.}(2007){Gerke}, {Newman}, {Faber}, {Cooper}, {Croton},
  {Davis}, {Willmer}, {Yan}, {Coil}, {Guhathakurta}, {Koo}, \&
  {Weiner}}]{gerke_etal:07}
{Gerke}, B.~F., {Newman}, J.~A., {Faber}, S.~M., {Cooper}, M.~C., {Croton},
  D.~J., {Davis}, M., {Willmer}, C.~N.~A., {Yan}, R., {Coil}, A.~L.,
  {Guhathakurta}, P., {Koo}, D.~C., \& {Weiner}, B.~J. 2007, \mnras, 376, 1425

\bibitem[{{Gunn} \& {Gott}(1972)}]{gunn_gott:72}
{Gunn}, J.~E. \& {Gott}, J.~R.~I. 1972, \apj, 176, 1

\bibitem[{{Hamilton}(1998)}]{hamilton:98}
{Hamilton}, A.~J.~S. 1998, in Astrophysics and Space Science Library, Vol. 231,
  The Evolving Universe, ed. D.~{Hamilton}, 185--+

\bibitem[{{Hansen} {et~al.}(2009){Hansen}, {Sheldon}, {Wechsler}, \&
  {Koester}}]{hansen_etal:09}
{Hansen}, S.~M., {Sheldon}, E.~S., {Wechsler}, R.~H., \& {Koester}, B.~P. 2009,
  \apj, 699, 1333

\bibitem[{{Harker} {et~al.}(2007){Harker}, {Cole}, \&
  {Jenkins}}]{harker_etal:07}
{Harker}, G., {Cole}, S., \& {Jenkins}, A. 2007, \mnras, 382, 1503

\bibitem[{{Hawkins} {et~al.}(2003){Hawkins}, {Maddox}, {Cole}, {Lahav},
  {Madgwick}, {Norberg}, {Peacock}, {Baldry}, {Baugh}, {Bland-Hawthorn},
  {Bridges}, {Cannon}, {Colless}, {Collins}, {Couch}, {Dalton}, {De Propris},
  {Driver}, {Efstathiou}, {Ellis}, {Frenk}, {Glazebrook}, {Jackson}, {Jones},
  {Lewis}, {Lumsden}, {Percival}, {Peterson}, {Sutherland}, \&
  {Taylor}}]{hawkins_etal:03}
{Hawkins}, E., {Maddox}, S., {Cole}, S., {Lahav}, O., {Madgwick}, D.~S.,
  {Norberg}, P., {Peacock}, J.~A., {Baldry}, I.~K., {Baugh}, C.~M.,
  {Bland-Hawthorn}, J., {Bridges}, T., {Cannon}, R., {Colless}, M., {Collins},
  C., {Couch}, W., {Dalton}, G., {De Propris}, R., {Driver}, S.~P.,
  {Efstathiou}, G., {Ellis}, R.~S., {Frenk}, C.~S., {Glazebrook}, K.,
  {Jackson}, C., {Jones}, B., {Lewis}, I., {Lumsden}, S., {Percival}, W.,
  {Peterson}, B.~A., {Sutherland}, W., \& {Taylor}, K. 2003, \mnras, 346, 78

\bibitem[{{Hopkins} {et~al.}(2008{\natexlab{a}}){Hopkins}, {Cox}, {Kere{\v s}},
  \& {Hernquist}}]{hopkins_etal:08b}
{Hopkins}, P.~F., {Cox}, T.~J., {Kere{\v s}}, D., \& {Hernquist}, L.
  2008{\natexlab{a}}, \apjs, 175, 390

\bibitem[{{Hopkins} {et~al.}(2008{\natexlab{b}}){Hopkins}, {Hernquist}, {Cox},
  \& {Kere{\v s}}}]{hopkins_etal:08a}
{Hopkins}, P.~F., {Hernquist}, L., {Cox}, T.~J., \& {Kere{\v s}}, D.
  2008{\natexlab{b}}, \apjs, 175, 356

\bibitem[{{Jannuzi} \& {Dey}(1999)}]{ndwfs}
{Jannuzi}, B.~T. \& {Dey}, A. 1999, in Astronomical Society of the Pacific
  Conference Series, Vol. 191, Photometric Redshifts and the Detection of High
  Redshift Galaxies, ed. R.~{Weymann}, L.~{Storrie-Lombardi}, M.~{Sawicki}, \&
  R.~{Brunner}, 111--+

\bibitem[{{Kaiser}(1987)}]{kaiser:87}
{Kaiser}, N. 1987, \mnras, 227, 1

\bibitem[{{Kauffmann} {et~al.}(2003{\natexlab{a}}){Kauffmann}, {Heckman},
  {White}, {Charlot}, {Tremonti}, {Brinchmann}, {Bruzual}, {Peng}, {Seibert},
  {Bernardi}, {Blanton}, {Brinkmann}, {Castander}, {Cs{\'a}bai}, {Fukugita},
  {Ivezic}, {Munn}, {Nichol}, {Padmanabhan}, {Thakar}, {Weinberg}, \&
  {York}}]{kauffmann_etal:03}
{Kauffmann}, G., {Heckman}, T.~M., {White}, S.~D.~M., {Charlot}, S.,
  {Tremonti}, C., {Brinchmann}, J., {Bruzual}, G., {Peng}, E.~W., {Seibert},
  M., {Bernardi}, M., {Blanton}, M., {Brinkmann}, J., {Castander}, F.,
  {Cs{\'a}bai}, I., {Fukugita}, M., {Ivezic}, Z., {Munn}, J.~A., {Nichol},
  R.~C., {Padmanabhan}, N., {Thakar}, A.~R., {Weinberg}, D.~H., \& {York}, D.
  2003{\natexlab{a}}, \mnras, 341, 33

\bibitem[{{Kauffmann} {et~al.}(2003{\natexlab{b}}){Kauffmann}, {Heckman},
  {White}, {Charlot}, {Tremonti}, {Peng}, {Seibert}, {Brinkmann}, {Nichol},
  {SubbaRao}, \& {York}}]{kauffmann_etal:03b}
{Kauffmann}, G., {Heckman}, T.~M., {White}, S.~D.~M., {Charlot}, S.,
  {Tremonti}, C., {Peng}, E.~W., {Seibert}, M., {Brinkmann}, J., {Nichol},
  R.~C., {SubbaRao}, M., \& {York}, D. 2003{\natexlab{b}}, \mnras, 341, 54

\bibitem[{{Kere{\v s}} {et~al.}(2008){Kere{\v s}}, {Katz}, {Fardal}, {Dave}, \&
  {Weinberg}}]{keres_etal:08}
{Kere{\v s}}, D., {Katz}, N., {Fardal}, M., {Dave}, R., \& {Weinberg}, D.~H.
  2008, \mnras, submitted (arXiv:0809.1430)

\bibitem[{{Kere{\v s}} {et~al.}(2005){Kere{\v s}}, {Katz}, {Weinberg}, \&
  {Dav{\'e}}}]{keres_etal:05}
{Kere{\v s}}, D., {Katz}, N., {Weinberg}, D.~H., \& {Dav{\'e}}, R. 2005,
  \mnras, 363, 2

\bibitem[{{Kimm} {et~al.}(2009){Kimm}, {Somerville}, {Yi}, {van den Bosch},
  {Salim}, {Fontanot}, {Monaco}, {Mo}, {Pasquali}, {Rich}, \&
  {Yang}}]{kimm_etal:09}
{Kimm}, T., {Somerville}, R.~S., {Yi}, S.~K., {van den Bosch}, F.~C., {Salim},
  S., {Fontanot}, F., {Monaco}, P., {Mo}, H., {Pasquali}, A., {Rich}, R.~M., \&
  {Yang}, X. 2009, \mnras, 394, 1131

\bibitem[{{Kochanek} {et~al.}(2003){Kochanek}, {White}, {Huchra}, {Macri},
  {Jarrett}, {Schneider}, \& {Mader}}]{kochanek_etal:03}
{Kochanek}, C.~S., {White}, M., {Huchra}, J., {Macri}, L., {Jarrett}, T.~H.,
  {Schneider}, S.~E., \& {Mader}, J. 2003, \apj, 585, 161

\bibitem[{{Koester} {et~al.}(2007){Koester}, {McKay}, {Annis}, {Wechsler},
  {Evrard}, {Bleem}, {Becker}, {Johnston}, {Sheldon}, {Nichol}, {Miller},
  {Scranton}, {Bahcall}, {Barentine}, {Brewington}, {Brinkmann}, {Harvanek},
  {Kleinman}, {Krzesinski}, {Long}, {Nitta}, {Schneider}, {Sneddin}, {Voges},
  \& {York}}]{koester_etal:07}
{Koester}, B.~P., {McKay}, T.~A., {Annis}, J., {Wechsler}, R.~H., {Evrard}, A.,
  {Bleem}, L., {Becker}, M., {Johnston}, D., {Sheldon}, E., {Nichol}, R.,
  {Miller}, C., {Scranton}, R., {Bahcall}, N., {Barentine}, J., {Brewington},
  H., {Brinkmann}, J., {Harvanek}, M., {Kleinman}, S., {Krzesinski}, J.,
  {Long}, D., {Nitta}, A., {Schneider}, D.~P., {Sneddin}, S., {Voges}, W., \&
  {York}, D. 2007, \apj, 660, 239

\bibitem[{{Kravtsov} {et~al.}(2004){Kravtsov}, {Berlind}, {Wechsler}, {Klypin},
  {Gottl{\" o}ber}, {Allgood}, \& {Primack}}]{kravtsov_etal:04}
{Kravtsov}, A.~V., {Berlind}, A.~A., {Wechsler}, R.~H., {Klypin}, A.~A.,
  {Gottl{\" o}ber}, S., {Allgood}, B., \& {Primack}, J.~R. 2004, \apj, 609, 35

\bibitem[{{Kriek} {et~al.}(2008){Kriek}, {van der Wel}, {van Dokkum}, {Franx},
  \& {Illingworth}}]{kriek_etal:08}
{Kriek}, M., {van der Wel}, A., {van Dokkum}, P.~G., {Franx}, M., \&
  {Illingworth}, G.~D. 2008, \apj, 682, 896

\bibitem[{{Kriek} {et~al.}(2006){Kriek}, {van Dokkum}, {Franx}, {Quadri},
  {Gawiser}, {Herrera}, {Illingworth}, {Labb{\'e}}, {Lira}, {Marchesini},
  {Rix}, {Rudnick}, {Taylor}, {Toft}, {Urry}, \& {Wuyts}}]{kriek_etal:06}
{Kriek}, M., {van Dokkum}, P.~G., {Franx}, M., {Quadri}, R., {Gawiser}, E.,
  {Herrera}, D., {Illingworth}, G.~D., {Labb{\'e}}, I., {Lira}, P.,
  {Marchesini}, D., {Rix}, H.-W., {Rudnick}, G., {Taylor}, E.~N., {Toft}, S.,
  {Urry}, C.~M., \& {Wuyts}, S. 2006, \apjl, 649, L71

\bibitem[{{Labb{\'e}} {et~al.}(2005){Labb{\'e}}, {Huang}, {Franx}, {Rudnick},
  {Barmby}, {Daddi}, {van Dokkum}, {Fazio}, {Schreiber}, {Moorwood}, {Rix},
  {R{\"o}ttgering}, {Trujillo}, \& {van der Werf}}]{labbe_etal:05}
{Labb{\'e}}, I., {Huang}, J., {Franx}, M., {Rudnick}, G., {Barmby}, P.,
  {Daddi}, E., {van Dokkum}, P.~G., {Fazio}, G.~G., {Schreiber}, N.~M.~F.,
  {Moorwood}, A.~F.~M., {Rix}, H.-W., {R{\"o}ttgering}, H., {Trujillo}, I., \&
  {van der Werf}, P. 2005, \apjl, 624, L81

\bibitem[{{Landy} \& {Szalay}(1993)}]{landy_szalay:93}
{Landy}, S.~D. \& {Szalay}, A.~S. 1993, \apj, 412, 64

\bibitem[{{Li} {et~al.}(2006){Li}, {Kauffmann}, {Jing}, {White}, {B{\"o}rner},
  \& {Cheng}}]{li_etal:06}
{Li}, C., {Kauffmann}, G., {Jing}, Y.~P., {White}, S.~D.~M., {B{\"o}rner}, G.,
  \& {Cheng}, F.~Z. 2006, \mnras, 368, 21

\bibitem[{{Lin} {et~al.}(2004){Lin}, {Mohr}, \& {Stanford}}]{lin_etal:04}
{Lin}, Y.-T., {Mohr}, J.~J., \& {Stanford}, S.~A. 2004, \apj, 610, 745

\bibitem[{{Madgwick} {et~al.}(2003){Madgwick}, {Somerville}, {Lahav}, \&
  {Ellis}}]{madgwick_etal:03}
{Madgwick}, D.~S., {Somerville}, R., {Lahav}, O., \& {Ellis}, R. 2003, \mnras,
  343, 871

\bibitem[{{Magliocchetti} \& {Porciani}(2003)}]{magliocchetti_porciani:03}
{Magliocchetti}, M. \& {Porciani}, C. 2003, \mnras, 346, 186

\bibitem[{{Maller}(2008)}]{maller:08}
{Maller}, A.~H. 2008, in Astronomical Society of the Pacific Conference Series,
  Vol. 396, Astronomical Society of the Pacific Conference Series, ed. J.~G.
  {Funes} \& E.~M. {Corsini}, 251--+

\bibitem[{{Maller} {et~al.}(2009){Maller}, {Berlind}, {Blanton}, \&
  {Hogg}}]{maller_etal:09}
{Maller}, A.~H., {Berlind}, A.~A., {Blanton}, M.~R., \& {Hogg}, D.~W. 2009,
  \apj, 691, 394

\bibitem[{{Marchesini} {et~al.}(2007){Marchesini}, {van Dokkum}, {Quadri},
  {Rudnick}, {Franx}, {Lira}, {Wuyts}, {Gawiser}, {Christlein}, \&
  {Toft}}]{marchesini_etal:07}
{Marchesini}, D., {van Dokkum}, P., {Quadri}, R., {Rudnick}, G., {Franx}, M.,
  {Lira}, P., {Wuyts}, S., {Gawiser}, E., {Christlein}, D., \& {Toft}, S. 2007,
  \apj, 656, 42

\bibitem[{{Mar{\'{\i}}n} {et~al.}(2008){Mar{\'{\i}}n}, {Wechsler}, {Frieman},
  \& {Nichol}}]{marin_etal:08}
{Mar{\'{\i}}n}, F.~A., {Wechsler}, R.~H., {Frieman}, J.~A., \& {Nichol}, R.~C.
  2008, \apj, 672, 849

\bibitem[{Mihos \& Hernquist(1996)}]{mihos_hernquist:96}
Mihos, J. \& Hernquist, L. 1996, \apj, 464, 641

\bibitem[{{More} {et~al.}(2009){More}, {van den Bosch}, {Cacciato}, {Mo},
  {Yang}, \& {Li}}]{more_etal:09}
{More}, S., {van den Bosch}, F.~C., {Cacciato}, M., {Mo}, H.~J., {Yang}, X., \&
  {Li}, R. 2009, \mnras, 392, 801

\bibitem[{{Moster} {et~al.}(2009){Moster}, {Somerville}, {Maulbetsch}, {van den
  Bosch}, {Maccio'}, {Naab}, \& {Oser}}]{moster_etal:09}
{Moster}, B.~P., {Somerville}, R.~S., {Maulbetsch}, C., {van den Bosch}, F.~C.,
  {Maccio'}, A.~V., {Naab}, T., \& {Oser}, L. 2009, \apj, submitted,
  ArXiv:0903.4682

\bibitem[{{Nagai} \& {Kravtsov}(2005)}]{nagai_kravtsov:05}
{Nagai}, D. \& {Kravtsov}, A.~V. 2005, \apj, 618, 557

\bibitem[{{Negroponte} \& {White}(1983)}]{negroponte_white:83}
{Negroponte}, J. \& {White}, S.~D.~M. 1983, \mnras, 205, 1009

\bibitem[{{Noeske} {et~al.}(2007){Noeske}, {Faber}, {Weiner}, {Koo}, {Primack},
  {Dekel}, {Papovich}, {Conselice}, {Le Floc'h}, {Rieke}, {Coil}, {Lotz},
  {Somerville}, \& {Bundy}}]{noeske_etal:07a}
{Noeske}, K.~G., {Faber}, S.~M., {Weiner}, B.~J., {Koo}, D.~C., {Primack},
  J.~R., {Dekel}, A., {Papovich}, C., {Conselice}, C.~J., {Le Floc'h}, E.,
  {Rieke}, G.~H., {Coil}, A.~L., {Lotz}, J.~M., {Somerville}, R.~S., \&
  {Bundy}, K. 2007, \apjl, 660, L47

\bibitem[{{Norberg} {et~al.}(2002){Norberg}, {Baugh}, {Hawkins}, {Maddox},
  {Madgwick}, {Lahav}, {Cole}, {Frenk}, {Baldry}, {Bland-Hawthorn}, {Bridges},
  {Cannon}, {Colless}, {Collins}, {Couch}, {Dalton}, {De Propris}, {Driver},
  {Efstathiou}, {Ellis}, {Glazebrook}, {Jackson}, {Lewis}, {Lumsden},
  {Peacock}, {Peterson}, {Sutherland}, \& {Taylor}}]{norberg_etal:02}
{Norberg}, P., {Baugh}, C.~M., {Hawkins}, E., {Maddox}, S., {Madgwick}, D.,
  {Lahav}, O., {Cole}, S., {Frenk}, C.~S., {Baldry}, I., {Bland-Hawthorn}, J.,
  {Bridges}, T., {Cannon}, R., {Colless}, M., {Collins}, C., {Couch}, W.,
  {Dalton}, G., {De Propris}, R., {Driver}, S.~P., {Efstathiou}, G., {Ellis},
  R.~S., {Glazebrook}, K., {Jackson}, C., {Lewis}, I., {Lumsden}, S.,
  {Peacock}, J.~A., {Peterson}, B.~A., {Sutherland}, W., \& {Taylor}, K. 2002,
  \mnras, 332, 827

\bibitem[{{Papovich} {et~al.}(2006){Papovich}, {Moustakas}, {Dickinson}, {Le
  Floc'h}, {Rieke}, {Daddi}, {Alexander}, {Bauer}, {Brandt}, {Dahlen}, {Egami},
  {Eisenhardt}, {Elbaz}, {Ferguson}, {Giavalisco}, {Lucas}, {Mobasher},
  {P{\'e}rez-Gonz{\'a}lez}, {Stutz}, {Rieke}, \& {Yan}}]{papovich_etal:06}
{Papovich}, C., {Moustakas}, L.~A., {Dickinson}, M., {Le Floc'h}, E., {Rieke},
  G.~H., {Daddi}, E., {Alexander}, D.~M., {Bauer}, F., {Brandt}, W.~N.,
  {Dahlen}, T., {Egami}, E., {Eisenhardt}, P., {Elbaz}, D., {Ferguson}, H.~C.,
  {Giavalisco}, M., {Lucas}, R.~A., {Mobasher}, B., {P{\'e}rez-Gonz{\'a}lez},
  P.~G., {Stutz}, A., {Rieke}, M.~J., \& {Yan}, H. 2006, \apj, 640, 92

\bibitem[{{Peacock} \& {Smith}(2000)}]{peacock_smith:00}
{Peacock}, J.~A. \& {Smith}, R.~E. 2000, \mnras, 318, 1144

\bibitem[{{Phleps} {et~al.}(2006){Phleps}, {Peacock}, {Meisenheimer}, \&
  {Wolf}}]{phleps_etal:06}
{Phleps}, S., {Peacock}, J.~A., {Meisenheimer}, K., \& {Wolf}, C. 2006, \aap,
  457, 145

\bibitem[{{Quadri} {et~al.}(2007){Quadri}, {van Dokkum}, {Gawiser}, {Franx},
  {Marchesini}, {Lira}, {Rudnick}, {Herrera}, {Maza}, {Kriek}, {Labb{\'e}}, \&
  {Francke}}]{quadri_etal:07}
{Quadri}, R., {van Dokkum}, P., {Gawiser}, E., {Franx}, M., {Marchesini}, D.,
  {Lira}, P., {Rudnick}, G., {Herrera}, D., {Maza}, J., {Kriek}, M.,
  {Labb{\'e}}, I., \& {Francke}, H. 2007, \apj, 654, 138

\bibitem[{{Quadri} {et~al.}(2008){Quadri}, {Williams}, {Lee}, {Franx}, {van
  Dokkum}, \& {Brammer}}]{quadri_etal:08}
{Quadri}, R.~F., {Williams}, R.~J., {Lee}, K.-S., {Franx}, M., {van Dokkum},
  P., \& {Brammer}, G.~B. 2008, \apjl, 685, L1

\bibitem[{{Scoccimarro}(2004)}]{roman:04}
{Scoccimarro}, R. 2004, \prd, 70, 083007

\bibitem[{{Scoccimarro} {et~al.}(2001){Scoccimarro}, {Sheth}, {Hui}, \&
  {Jain}}]{roman_etal:01}
{Scoccimarro}, R., {Sheth}, R.~K., {Hui}, L., \& {Jain}, B. 2001, \apj, 546, 20

\bibitem[{{Scranton}(2003)}]{scranton:03}
{Scranton}, R. 2003, \mnras, 339, 410

\bibitem[{{Seljak}(2000)}]{seljak:00}
{Seljak}, U. 2000, \mnras, 318, 203

\bibitem[{{Seljak} {et~al.}(2009){Seljak}, {Hamaus}, \&
  {Desjacques}}]{seljak_etal:09}
{Seljak}, U., {Hamaus}, N., \& {Desjacques}, V. 2009, ArXiv e-prints

\bibitem[{{Skibba} \& {Sheth}(2009)}]{skibba_sheth:09}
{Skibba}, R.~A. \& {Sheth}, R.~K. 2009, \mnras, 392, 1080

\bibitem[{{Somerville} {et~al.}(2008){Somerville}, {Hopkins}, {Cox},
  {Robertson}, \& {Hernquist}}]{somerville_etal:08}
{Somerville}, R.~S., {Hopkins}, P.~F., {Cox}, T.~J., {Robertson}, B.~E., \&
  {Hernquist}, L. 2008, \mnras, 1241

\bibitem[{{Springel}(2000)}]{springel:00}
{Springel}, V. 2000, \mnras, 312, 859

\bibitem[{{Strateva} {et~al.}(2001){Strateva}, {Ivezi{\'c}}, {Knapp},
  {Narayanan}, {Strauss}, {Gunn}, {Lupton}, {Schlegel}, {Bahcall}, {Brinkmann},
  {Brunner}, {Budav{\'a}ri}, {Csabai}, {Castander}, {Doi}, {Fukugita}, {Gy{\H
  o}ry}, {Hamabe}, {Hennessy}, {Ichikawa}, {Kunszt}, {Lamb}, {McKay},
  {Okamura}, {Racusin}, {Sekiguchi}, {Schneider}, {Shimasaku}, \&
  {York}}]{strateva_etal:01}
{Strateva}, I., {Ivezi{\'c}}, {\v Z}., {Knapp}, G.~R., {Narayanan}, V.~K.,
  {Strauss}, M.~A., {Gunn}, J.~E., {Lupton}, R.~H., {Schlegel}, D., {Bahcall},
  N.~A., {Brinkmann}, J., {Brunner}, R.~J., {Budav{\'a}ri}, T., {Csabai}, I.,
  {Castander}, F.~J., {Doi}, M., {Fukugita}, M., {Gy{\H o}ry}, Z., {Hamabe},
  M., {Hennessy}, G., {Ichikawa}, T., {Kunszt}, P.~Z., {Lamb}, D.~Q., {McKay},
  T.~A., {Okamura}, S., {Racusin}, J., {Sekiguchi}, M., {Schneider}, D.~P.,
  {Shimasaku}, K., \& {York}, D. 2001, \aj, 122, 1861

\bibitem[{{Tinker} {et~al.}(2008{\natexlab{a}}){Tinker}, {Kravtsov}, {Klypin},
  {Abazajian}, {Warren}, {Yepes}, {Gottl{\"o}ber}, \&
  {Holz}}]{tinker_etal:08_mf}
{Tinker}, J., {Kravtsov}, A.~V., {Klypin}, A., {Abazajian}, K., {Warren}, M.,
  {Yepes}, G., {Gottl{\"o}ber}, S., \& {Holz}, D.~E. 2008{\natexlab{a}}, \apj,
  688, 709

\bibitem[{{Tinker} {et~al.}(2008{\natexlab{b}}){Tinker}, {Conroy}, {Norberg},
  {Patiri}, {Weinberg}, \& {Warren}}]{tinker_etal:08_voids}
{Tinker}, J.~L., {Conroy}, C., {Norberg}, P., {Patiri}, S.~G., {Weinberg},
  D.~H., \& {Warren}, M.~S. 2008{\natexlab{b}}, \apj, 686, 53

\bibitem[{{Tinker} {et~al.}(2007){Tinker}, {Norberg}, {Weinberg}, \&
  {Warren}}]{tinker_etal:07_pvd}
{Tinker}, J.~L., {Norberg}, P., {Weinberg}, D.~H., \& {Warren}, M.~S. 2007,
  \apj, 659, 877

\bibitem[{{Tinker} {et~al.}(2009{\natexlab{a}}){Tinker}, {Wechsler}, \&
  {Zheng}}]{tinker_etal:09_drg}
{Tinker}, J.~L., {Wechsler}, R.~H., \& {Zheng}, Z. 2009{\natexlab{a}}, \apj,
  (submitted), arXiv:0902.1748

\bibitem[{{Tinker} {et~al.}(2006){Tinker}, {Weinberg}, \&
  {Warren}}]{tinker_etal:06_voids}
{Tinker}, J.~L., {Weinberg}, D.~H., \& {Warren}, M.~S. 2006, \apj, 647, 737

\bibitem[{{Tinker} {et~al.}(2005){Tinker}, {Weinberg}, {Zheng}, \&
  {Zehavi}}]{tinker_etal:05}
{Tinker}, J.~L., {Weinberg}, D.~H., {Zheng}, Z., \& {Zehavi}, I. 2005, \apj,
  631, 41

\bibitem[{{Tinker} {et~al.}(2009{\natexlab{b}})}]{tinker_etal:09_bias}
{Tinker}, J.~T. {et~al.} 2009{\natexlab{b}}, in preparation

\bibitem[{{Toomre} \& {Toomre}(1972)}]{toomre_toomre:72}
{Toomre}, A. \& {Toomre}, J. 1972, \apj, 178, 623

\bibitem[{{Vale} \& {Ostriker}(2006)}]{vale_ostriker:06}
{Vale}, A. \& {Ostriker}, J.~P. 2006, \mnras, 371, 1173

\bibitem[{{van den Bosch} {et~al.}(2008){van den Bosch}, {Aquino}, {Yang},
  {Mo}, {Pasquali}, {McIntosh}, {Weinmann}, \& {Kang}}]{vdb_etal:08}
{van den Bosch}, F.~C., {Aquino}, D., {Yang}, X., {Mo}, H.~J., {Pasquali}, A.,
  {McIntosh}, D.~H., {Weinmann}, S.~M., \& {Kang}, X. 2008, \mnras, 387, 79

\bibitem[{{van den Bosch} {et~al.}(2007){van den Bosch}, {Yang}, {Mo},
  {Weinmann}, {Macci{\`o}}, {More}, {Cacciato}, {Skibba}, \&
  {Kang}}]{vdb_etal:07}
{van den Bosch}, F.~C., {Yang}, X., {Mo}, H.~J., {Weinmann}, S.~M.,
  {Macci{\`o}}, A.~V., {More}, S., {Cacciato}, M., {Skibba}, R., \& {Kang}, X.
  2007, \mnras, 376, 841

\bibitem[{{Wang} {et~al.}(2006){Wang}, {Li}, {Kauffmann}, \& {De
  Lucia}}]{wang_etal:06}
{Wang}, L., {Li}, C., {Kauffmann}, G., \& {De Lucia}, G. 2006, \mnras, 371, 537

\bibitem[{{Wang} {et~al.}(2007){Wang}, {Li}, {Kauffmann}, \& {De
  Lucia}}]{wang_etal:07}
---. 2007, \mnras, 377, 1419

\bibitem[{{Weinmann} {et~al.}(2006){Weinmann}, {van den Bosch}, {Yang}, \&
  {Mo}}]{weinmann_etal:06}
{Weinmann}, S.~M., {van den Bosch}, F.~C., {Yang}, X., \& {Mo}, H.~J. 2006,
  \mnras, 366, 2

\bibitem[{{Wetzel} \& {White}(2009)}]{wetzel_white:09}
{Wetzel}, A. \& {White}, M. 2009, \mnras, submitted, ArXiv:0907.0702

\bibitem[{{Wetzel} {et~al.}(2009){Wetzel}, {Cohn}, \&
  {White}}]{wetzel_etal:09b}
{Wetzel}, A.~R., {Cohn}, J.~D., \& {White}, M. 2009, \mnras, 395, 1376

\bibitem[{{White}(2002)}]{white:02}
{White}, M. 2002, \apjs, 143, 241

\bibitem[{{White} {et~al.}(2001){White}, {Hernquist}, \&
  {Springel}}]{white_etal:01}
{White}, M., {Hernquist}, L., \& {Springel}, V. 2001, \apjl, 550, L129

\bibitem[{{Williams} {et~al.}(2009){Williams}, {Quadri}, {Franx}, {van Dokkum},
  \& {Labb{\'e}}}]{williams_etal:09}
{Williams}, R.~J., {Quadri}, R.~F., {Franx}, M., {van Dokkum}, P., \&
  {Labb{\'e}}, I. 2009, \apj, 691, 1879

\bibitem[{{Willmer} {et~al.}(2006){Willmer}, {Faber}, {Koo}, {Weiner},
  {Newman}, {Coil}, {Connolly}, {Conroy}, {Cooper}, {Davis}, {Finkbeiner},
  {Gerke}, {Guhathakurta}, {Harker}, {Kaiser}, {Kassin}, {Konidaris}, {Lin},
  {Luppino}, {Madgwick}, {Noeske}, {Phillips}, \& {Yan}}]{willmer_etal:06}
{Willmer}, C.~N.~A., {Faber}, S.~M., {Koo}, D.~C., {Weiner}, B.~J., {Newman},
  J.~A., {Coil}, A.~L., {Connolly}, A.~J., {Conroy}, C., {Cooper}, M.~C.,
  {Davis}, M., {Finkbeiner}, D.~P., {Gerke}, B.~F., {Guhathakurta}, P.,
  {Harker}, J., {Kaiser}, N., {Kassin}, S., {Konidaris}, N.~P., {Lin}, L.,
  {Luppino}, G., {Madgwick}, D.~S., {Noeske}, K.~G., {Phillips}, A.~C., \&
  {Yan}, R. 2006, \apj, 647, 853

\bibitem[{{Wolf} {et~al.}(2004){Wolf}, {Meisenheimer}, {Kleinheinrich},
  {Borch}, {Dye}, {Gray}, {Wisotzki}, {Bell}, {Rix}, {Cimatti}, {Hasinger}, \&
  {Szokoly}}]{wolf_etal:04}
{Wolf}, C., {Meisenheimer}, K., {Kleinheinrich}, M., {Borch}, A., {Dye}, S.,
  {Gray}, M., {Wisotzki}, L., {Bell}, E.~F., {Rix}, H.-W., {Cimatti}, A.,
  {Hasinger}, G., \& {Szokoly}, G. 2004, \aap, 421, 913

\bibitem[{{Wolf} {et~al.}(2003){Wolf}, {Meisenheimer}, {Rix}, {Borch}, {Dye},
  \& {Kleinheinrich}}]{wolf_etal:03}
{Wolf}, C., {Meisenheimer}, K., {Rix}, H.-W., {Borch}, A., {Dye}, S., \&
  {Kleinheinrich}, M. 2003, \aap, 401, 73

\bibitem[{{Yang} {et~al.}(2003){Yang}, {Mo}, \& {van den Bosch}}]{yang_etal:03}
{Yang}, X., {Mo}, H.~J., \& {van den Bosch}, F.~C. 2003, \mnras, 339, 1057

\bibitem[{{Yang} {et~al.}(2008){Yang}, {Mo}, \& {van den Bosch}}]{yang_etal:08}
---. 2008, \apj, 676, 248

\bibitem[{{Zehavi} {et~al.}(2005){Zehavi}, {Zheng}, {Weinberg}, {Frieman},
  {Berlind}, {Blanton}, {Scoccimarro}, {Sheth}, {Strauss}, {Kayo}, {Suto},
  {Fukugita}, {Nakamura}, {Bahcall}, {Brinkmann}, {Gunn}, {Hennessy},
  {Ivezi{\'c}}, {Knapp}, {Loveday}, {Meiksin}, {Schlegel}, {Schneider},
  {Szapudi}, {Tegmark}, {Vogeley}, \& {York}}]{zehavi_etal:05}
{Zehavi}, I., {Zheng}, Z., {Weinberg}, D.~H., {Frieman}, J.~A., {Berlind},
  A.~A., {Blanton}, M.~R., {Scoccimarro}, R., {Sheth}, R.~K., {Strauss}, M.~A.,
  {Kayo}, I., {Suto}, Y., {Fukugita}, M., {Nakamura}, O., {Bahcall}, N.~A.,
  {Brinkmann}, J., {Gunn}, J.~E., {Hennessy}, G.~S., {Ivezi{\'c}}, {\v Z}.,
  {Knapp}, G.~R., {Loveday}, J., {Meiksin}, A., {Schlegel}, D.~J., {Schneider},
  D.~P., {Szapudi}, I., {Tegmark}, M., {Vogeley}, M.~S., \& {York}, D.~G. 2005,
  \apj, 630, 1

\bibitem[{{Zhao} {et~al.}(2008){Zhao}, {Jing}, {Mo}, \&
  {Boerner}}]{zhao_etal:08}
{Zhao}, D.~H., {Jing}, Y.~P., {Mo}, H.~J., \& {Boerner}, G. 2008, ArXiv
  e-prints

\bibitem[{{Zheng}(2004)}]{zheng:04}
{Zheng}, Z. 2004, \apj, 610, 61

\bibitem[{{Zheng} {et~al.}(2005){Zheng}, {Berlind}, {Weinberg}, {Benson},
  {Baugh}, {Cole}, {Dav{\'e}}, {Frenk}, {Katz}, \& {Lacey}}]{zheng_etal:05}
{Zheng}, Z., {Berlind}, A.~A., {Weinberg}, D.~H., {Benson}, A.~J., {Baugh},
  C.~M., {Cole}, S., {Dav{\'e}}, R., {Frenk}, C.~S., {Katz}, N., \& {Lacey},
  C.~G. 2005, \apj, 633, 791

\bibitem[{{Zheng} {et~al.}(2007){Zheng}, {Coil}, \& {Zehavi}}]{zheng_etal:07}
{Zheng}, Z., {Coil}, A.~L., \& {Zehavi}, I. 2007, \apj, 667, 760

\bibitem[{{Zheng} {et~al.}(2008){Zheng}, {Zehavi}, {Eisenstein}, {Weinberg}, \&
  {Jing}}]{zheng_etal:08}
{Zheng}, Z., {Zehavi}, I., {Eisenstein}, D.~J., {Weinberg}, D.~H., \& {Jing},
  Y. 2008, \apj, submitted, arXiv:0809.1868 [astro-ph]

\end{thebibliography}

\end{document}